%% file: modular_numerics.tex
\documentclass[11pt,a4paper]{article}
\usepackage{jheppub}

\usepackage{dsfont}

\usepackage{amsmath}
\usepackage{amsfonts}
\usepackage{slashed}
\usepackage{graphicx}
\usepackage{mathrsfs}
\usepackage{tikz}
\usepackage[numbers,sort&compress]{natbib}
\usepackage{subcaption}

\makeatletter
\g@addto@macro\bfseries{\boldmath}
\makeatother

\newcommand{\eps}{\epsilon}
\newcommand{\ord}{\begin{cal}O\end{cal}}

\def\beq{\begin{equation}}
\def\eeq{\end{equation}}
\def\bsp#1\esp{\begin{split}#1\end{split}}

\makeatletter
\newcommand{\IEIF}{%
  \def\@IEIFsep{(}%
  I_F\@IEIFi
}
\newcommand\@IEIFi{\@ifnextchar\stopIEIF{\@IEIFend}{\@IEIFii}}
\newcommand\@IEIFii[4]{%
  \big\@IEIFsep
  \begin{smallmatrix}
    #1 & #2 \\
    #3 & #4
  \end{smallmatrix}
  \def\@IEIFsep{|}
  \@IEIFi
}
\newcommand\@IEIFend[2]{%
  ; #2 \bigr)
}
\makeatother

\newcommand{\cE}{\begin{cal}E\end{cal}}
\newcommand{\cF}{\begin{cal}F\end{cal}}

\newcommand{\cM}{\begin{cal}M\end{cal}}

\newcommand{\cS}{\begin{cal}S\end{cal}}
\newcommand{\cT}{\begin{cal}T\end{cal}}

\newcommand{\cW}{\begin{cal}W\end{cal}}

\newcommand{\ddl}{\mathfrak{D}^d\ell}

\def \sha{{\,\amalg\hskip -3.8pt\amalg\,}}

\newcommand{\IMFo}[2]{\mathcal{I}\hskip-0.08cm\left(\begin{smallmatrix} #1\end{smallmatrix};#2\right)}

\newcommand{\mat}[2]{\left(\begin{smallmatrix}#1 \\ #2\end{smallmatrix}\right)}

\newcommand{\sh}{\sha}

\DeclareMathOperator{\SL}{SL}
\DeclareMathOperator{\K}{K}
\DeclareMathOperator{\E}{E}

\newcommand*{\eisfont}{\fontfamily{phv}\selectfont}
\DeclareTextFontCommand{\texteisfont}{\eisfont}
\newcommand{\eisa}[0]{\small{\texteisfont{a}}}
\newcommand{\eisb}[0]{\small{\texteisfont{b}}}

\allowdisplaybreaks

\title{Algorithms and tools for iterated Eisenstein integrals}

\author[a]{Claude Duhr}
\author[b]{Lorenzo Tancredi}

\affiliation[a]{Theoretical Physics Department, CERN, Geneva, Switzerland} 
\affiliation[b]{Rudolf Peierls Centre for Theoretical Physics, Clarendon Laboratory, Parks Road, Oxford OX1 3PU, UK}

\emailAdd{claude.duhr@cern.ch}
\emailAdd{lorenzo.tancredi@physics.ox.ac.uk}

\abstract{We present algorithms to work with iterated Eisenstein integrals that have recently appeared in the computation of multi-loop 
Feynman integrals. These algorithms allow one to analytically continue these integrals to all regions of the parameter space, and to obtain fast converging series representations in each region. We illustrate our approach on the examples of hypergeometric functions that evaluate to iterated Eisenstein integrals as well as the well-known sunrise graph.
}


\begin{document}

\begin{flushright}
\preprint{CERN-TH-2019-209, OUTP-19-14P}
\end{flushright}

\maketitle

\catcode`\@=11
\font\manfnt=manfnt
\def\Watchout{\@ifnextchar [{\W@tchout}{\W@tchout[1]}}
\def\W@tchout[#1]{{\manfnt\@tempcnta#1\relax%
  \@whilenum\@tempcnta>\z@\do{%
    \char"7F\hskip 0.3em\advance\@tempcnta\m@ne}}}
\let\foo\W@tchout
\def\dubious{\@ifnextchar[{\@dubious}{\@dubious[1]}}
\let\enddubious\endlist
\def\@dubious[#1]{%
  \setbox\@tempboxa\hbox{\@W@tchout#1}
  \@tempdima\wd\@tempboxa
  \list{}{\leftmargin\@tempdima}\item[\hbox to 0pt{\hss\@W@tchout#1}]}
\def\@W@tchout#1{\W@tchout[#1]}
\catcode`\@=12

\input{introduction}

\input{modular_forms}

\input{ieis}

\input{mpls}
\input{regularisation}

\input{numerics}

\input{F21}

\input{sunrise}

\input{conclusion}

\section*{Acknowledgments}
We thank Nils Matthes for interesting discussions and Brenda Penante for collaboration during the early stages of this project. 
This research was supported by the ERC starting grant  637019 ``MathAm'' and by the Royal Society through the grant URF\textbackslash R1\textbackslash 191125.

\appendix 
\input{app_SL2Z}

\input{app_2f1_results}
\input{app_sun_results}

\bibliography{bib}

\end{document}

%% file: introduction.tex

\section{Introduction and motivation}
\label{sec:intro}

During the past ten years, the Large Hadron Collider (LHC) at CERN in Geneva has been 
providing us with 
an impressive amount of data on the dynamics of the fundamental constituents
of matter at unprecedented energies and with unprecedented precision.
Interpreting correctly this data against theoretical predictions in 
the Standard Model offers us a precious opportunity to discover 
signs of New Physics lurking at higher energy scales. 
The Standard Model is based on Quantum Field Theory (QFT), and in QFT the probability 
amplitude for a scattering process  
is encoded in so-called scattering amplitudes.
Scattering amplitudes are in turn usually computed perturbatively 
through an expansion in Feynman diagrams, whose
calculation can be reduced to the evaluation of increasingly complicated multi-loop
scalar Feynman integrals.
While alternative approaches exist which
attempt to avoid the combinatorial complexity introduced by an 
expansion in off-shell Feynman diagrams, eventually
one always falls back to the calculation of scalar multi-loop Feynman integrals. 
The latter can therefore be thought as irreducible building blocks for the computation of any physical observable.

As dictated by causality and unitarity in QFT, scattering amplitudes 
are typically complex functions with a highly non-trivial branch-cut structure,
which stems from those regions of the phase-space where virtual particles
running in the loops are allowed to go on-shell. 
The details of their analytic structure must be encoded in the 
special functions used to represent them, which is expected to become
prohibitively complicated with the increase of the complexity of the process and the perturbative
order considered.

In the past two decades important steps forward have been
achieved towards the classification of the special functions that appear in multi-loop calculations.
The first crucial step in this direction was the realisation of the importance of so-called
multiple polylogarithms (MPLs) for scattering amplitudes
with (mainly) massless internal and external particles. MPLs are the natural generalisation
of the logarithm, and they can be defined (loosely) as the class of special functions required to 
perform all iterated integrals of rational functions with poles on the complex plane.
While MPLs had been known to the mathematical 
community for over a century~\cite{Nielsen, Kummer}, their re-discovery in the high-energy
physics community~\cite{Remiddi:1999ew,Gehrmann:2000zt} contributed to initiate a new spurt of interest in these functions. This has 
ultimately led to a much deeper understanding of their analytic and algebraic properties~\cite{Goncharov:2010jf,GoncharovMixedTate,Duhr:2011zq,Duhr:2012fh}, 
which is important in pehnomenological applications.
Indeed, these developments were crucial to make
a multitude of precision calculations in the Standard Model possible and, in this way, to  provide
a theoretical interpretation of the most recent results obtained at the LHC. 
Scattering amplitudes are typically only the first step towards 
a complete physical prediction. For example, at the LHC the calculation of differential distributions
for interesting observables requires evaluating complicated multi-dimensional 
phase-space integrals over the scattering amplitudes, which can in general only be
performed numerically by  Monte Carlo techniques. The latter typically require being able to
evaluate numerically the scattering amplitudes (hundreds of) thousands of times, in
different regions of the relevant phase-space, typically crossing multiple
branch cuts. For this reason, the importance of being able
to evaluate numerically fast and with high-precision the special functions required and 
to properly handle their branch cuts, cannot be overestimated for realistic physical applications.

Thanks to the developments hinted to above, today we can claim that both goals are achievable
as long as the scattering amplitudes
can be expressed in terms of MPLs only. Rather generic codes have been
published which can evaluate numerically MPLs for (in principle) arbitrary complex
arguments~\cite{Gehrmann:2001pz,Gehrmann:2001jv,Buehler:2011ev,Vollinga:2004sn,Frellesvig:2016ske,Ablinger:2018sat}. Still, in cases which are complicated enough, a brute force application of these 
algorithms appears unfeasible, as the numerical evaluation for arbitrary phase-space points
across multiple branch-cuts tends to become too slow and inefficient for their use in Monte Carlo
integration codes. Fortunately, this problem can be greatly alleviated by properly
manipulating the relevant MPLs and re-expressing them in terms of new combinations of
MPLs which can be more easily evaluated numerically in the region of interest. To achieve this,
one typically uses the knowledge of the algebraic and analytical properties of MPLs in order
to express MPLs whose numerical evaluation would require crossing complicated branch cuts, 
in terms of MPLs which are free of those branch cuts in the region considered and whose numerical
values can be obtained by fast converging series expansions. 
While it cannot be proved that this strategy will always be successful, its potential has been
shown in many examples, most notably in the evaluation of the two-loop QCD corrections
to the productions of pairs of vector bosons  at the LHC~\cite{Gehrmann:2014bfa,Gehrmann:2015ora,vonManteuffel:2015msa}. 
It is worth stressing that this step was essential to make the a large number of phenomenological studies at the LHC possible.

In spite of these impressive results, this is not the end of the story. As it is by now very well known, starting at 
the two-loop order MPLs are not enough to express all interesting scattering amplitudes.
The first occurrence of classes of iterated integrals beyond MPLs can be traced back to a paper by A. Sabry
in 1962, where he discovered new functions of elliptic type in the calculation of the two-loop corrections
to the electron propagator in QED with massive electrons~\cite{Sabry}. 
Renewed interest in these functions
arose when it was realised that they appear in a large number of calculations relevant for collider physics,
most notably in the two-loop correction to $t\bar{t}$ production at hadron colliders~\cite{Czakon:2013goa}. 
The simplest occurrence of these new functions
was identified in the so-called massive two-loop sunrise graph, whose computation has since then received a lot of 
attention from different sides of the high-energy physics community~\cite{Broadhurst:1987ei,Bauberger:1994by,Bauberger:1994hx,MullerStach:2011ru,Laporta:2004rb,Caffo:2008aw,Remiddi:2016gno,Broedel:2017siw,Remiddi:2013joa,Adams:2015ydq,Caffo:1998du,Adams:2014vja,Adams:2013kgc,Adams:2013nia,Adams:2015gva,Bloch:2016izu,Bloch:2013tra,Kniehl:2005bc,Bogner:2019lfa}.
Indeed, it has been recently realised that a natural generalisation of MPLs to include the new class of functions
that appear in the sunrise graph exists. These so-called elliptic multiple polylogarithms  (eMPLs) can be obtained
by considering iterated integrals of rational functions on a genus one complex surface, i.e. a torus, which
is well known to be mathematically equivalent to an elliptic curve.
Their description as iterated integrals on a complex torus had been formalised by mathematicians 
already in 2011~\cite{LevinRacinet,BrownLevin} and had later found 
application in the calculation of one-loop  scattering amplitudes
in string theory~\cite{Broedel:2014vla,Broedel:2018izr,Broedel:2015hia,Broedel:2017jdo}.
More recently, an alternative formulation of eMPLs naturally defined on an elliptic curved defined by a polynomial equation
instead of on a torus has been proposed~\cite{Broedel:2017kkb,Broedel:2017siw,Broedel:2018iwv}, which has made it possible to naturally compute
different previously out-of-reach multi-loop Feynman integrals~\cite{Broedel:2018qkq,Blumlein:2018jgc,Broedel:2019hyg}.
While eMPLs are a rather large class of functions that can encompass different kinds of physical problems, 
for the particular case of the two-loop sunrise graph,\footnote{And also of other similar problems depending on one single
independent variable.} a special subclass of eMPLs has been shown to be sufficient, i.e. the so-called
iterated Eisenstein integrals~\cite{Adams:2018ulb,Broedel:2018rwm,ZagierModular,Adams:2017ejb,ManinModular}.

In the past years there has been an impressive progress in understanding the analytical and algebraic properties of these new functions.
In spite of that, algorithms for the fast and precise numerical evaluation of iterated Eisenstein integrals and, 
more in general eMPLs, are still missing.\footnote{For specific instances of elliptic Feynman integrals fast numerical codes were achieved, see for example refs.~\cite{Remiddi:2016gno,Primo:2017ipr,vonManteuffel:2017hms,Bogner:2017vim,Frellesvig:2019byn}.} Without this step, their use in state-of-the-art physical
calculations will not become possible.
This paper aims to start closing this gap, at least for the restricted set of one-parameter Feynman integrals that can be expressed in terms of 
iterated Eisenstein integrals. Examples of such integrals involve the well-known two-loop equal-mass  sunrise and kite integrals~\cite{Adams:2017ejb,Broedel:2018iwv}, as well as the three-loop
banana graph~\cite{Broedel:2019kmn}. Indeed, we will show that for the case of iterated Eisenstein integrals, algorithms 
similar to those used to treat MPLs can be devised, which allow one to
efficiently evaluate complicated combinations of these functions for any values of the external 
kinematics of the problem considered. Similarly to what is done with MPLs, we will show that given a Feynman
integral expressed in terms of iterated Eisenstein integrals, irrespective of the kinematical
point we are interested in, it is always possible to rewrite it in terms of functions which can be numerically
evaluated by a fast converging series expansion, making their use suitable for realistic physical applications.
The rest of the paper is organised as follows: in Section~\ref{sec:modular_forms} we review the 
definition of modular forms in particular showing how to define a parity-invariant basis for the latter,
which is explicitly real or imaginary.
In Section~\ref{sec:ieis} we define 
iterated integrals of modular forms for this new parity-invariant basis and show how to use it to express 
a special class of hypergeometric functions.
We then have a short intermezzo in Section~\ref{sec:eMPLs_to_MPLs} to discuss under which assumptions it
is possible to express iterated Eisenstein integrals in terms of simpler MPLs. 
In Section~\ref{sec:modular_trafo} we then come to the central idea of this paper, which consists in an algorithm to 
perform modular transformations on iterated Eisenstein integrals in order to rewrite them in terms of combinations of iterated Eisenstein integrals
which can be more easily numerically evaluated. The details of the numerical evaluation are discussed
in Section~\ref{sec:numerics}, while explicit examples to classes of hypergeometric functions and to the sunrise integral
are provided in Sections~\ref{sec:2F1} and~\ref{sec:sunrise} respectively.
Finally, we draw our conclusions in Section~\ref{sec:conclusion}.

%% file: modular_forms.tex

\section{Modular forms and Eisenstein series}
\label{sec:modular_forms}

The purpose of this section is to review in detail the modular group $\SL(2,\mathbb{Z})$ and modular forms. Most of the material in this section is well known in the mathematics literature. We include it nevertheless because modular forms and their properties are the foundations on which subsequent sections are built.

\subsection{The modular group and its congruence subgroups} 
The group $\SL(2,\mathbb{Z})$ consists of all $2\times2$ matrices with integer entries and unit determinant. It is generated by the three elements
\beq\label{eq:SL2Z_generators}
S = S^{-1} = \mat{0 & -1}{1 & 0}\,,\qquad T = \mat{1&1}{0&1}\,, \qquad -\mathds{1} = \mat{-1&0}{0&-1}\,.
\eeq
In other words, every element of $\SL(2,\mathbb{Z})$ can be written as a finite product of $S$'s and/or $T$'s up to a sign. The decomposition of an element of $\SL(2,\mathbb{Z})$ into a product of generators can be performed algorithmically (see Appendix~\ref{app:SL2Z} for a review).

We are often not interested in the whole modular group $\SL(2,\mathbb{Z})$, but only in a certain subgroup thereof. Of particular interest are the so-called \emph{congruence subgroups}. A subgroup $\Gamma\subseteq \SL(2,\mathbb{Z})$ is called a congruence subgroup of level $N$ if it contains the principle congruence subgroup of level $N$, 
\beq\label{eq:Gamma(N)_def}
\Gamma(N) = \left\{\left(\begin{smallmatrix}a & b\\ c& d\end{smallmatrix}\right)\in \SL(2,\mathbb{Z}): a,d=1\!\!\! \mod N \textrm{ and } b,c=0\!\!\!\mod N\right\}\,.
\eeq
Particularly important examples of congruence subgroups are
\beq\bsp\label{eq:Gamma01(N)_def}
\Gamma_0(N) &\,= \left\{\left(\begin{smallmatrix}a & b\\ c& d\end{smallmatrix}\right)\in \SL(2,\mathbb{Z}): c=0\!\!\!\mod N\right\}\,,\\
\Gamma_1(N) &\,= \left\{\left(\begin{smallmatrix}a & b\\ c& d\end{smallmatrix}\right)\in \SL(2,\mathbb{Z}): a,d=1\!\!\! \mod N \textrm{ and } c=0\!\!\!\mod N\right\}\,.
\esp\eeq

For the rest of this section, $\Gamma$ will denote some congruence subgroup of $\SL(2,\mathbb{Z})$ (which may be $\SL(2,\mathbb{Z})=\Gamma(1)$ itself).
$\Gamma$ naturally acts on the upper half-plane $\mathbb{H}=\{\tau\in\mathbb{C}|\textrm{Im }\tau > 0\}$ via M\"obius transformations, 
\beq\label{eq:tau_trafo}
\gamma\cdot \tau = \frac{a\tau+b}{c\tau+d}\,,\qquad \gamma=\mat{a&b}{c&d}\in \Gamma\,.
\eeq
Note that $-\mathds{1}$ acts trivially on $\tau$, i.e., $(-\mathds{1})\cdot \tau=\tau$.
The action in eq.~\eqref{eq:tau_trafo} partitions the upper half-plane into distinct orbits. 
The space $Y_{\Gamma}={\mathbb{H}}/\Gamma$ of orbits is called the \emph{modular curve} for $\Gamma$. 
It can be useful to consider a connected domain in ${\mathbb{H}}$ whose points correspond to the distinct orbits in $Y_{\Gamma}$, called a \emph{fundamental domain} for $\Gamma$. A particularly important example of a fundamental domain is the one for the full modular group $\SL(2,\mathbb{Z})$, which is 
\beq\label{eq:fundamental_domain}
\!\!\cF = \{\tau\in{\mathbb{H}}:{-{1}/{2}}\le\textrm{Re }\tau<{1}/{2}\textrm{~and~}|\tau|>1\}\cup \{\tau\in{\mathbb{H}}:\textrm{Re }\tau \le0\textrm{~and~}|\tau|=1\}\,.
\eeq
In other words, for every $\tau\in{\mathbb{H}}$ there are unique $\gamma\in\SL(2,\mathbb{Z})$ and $\tau_0\in\cF$ such that $\tau=\gamma\cdot \tau_0$.

$\Gamma$ also acts on $\mathbb{Q}\cup\{i\infty\}$, and so it is natural to consider its action on the extended upper half-plane $\overline{\mathbb{H}}=\mathbb{H}\cup\mathbb{Q}\cup\{i\infty\}$.
It partitions $\mathbb{Q}\cup\{i\infty\}$ into a finite number of distinct equivalence classes, called the \emph{cusps} of $\Gamma$. For every $c_1,c_2\in\mathbb{Q}\cup\{i\infty\}$, there is a $\gamma\in\SL(2,\mathbb{Z})$ such that $c_2=\gamma\cdot c_1$, which implies that $\SL(2,\mathbb{Z})$ has only one cusp. The matrix $\gamma$ can be constructed explicitly: First, we note that it is sufficient to consider $c_1=i\infty$. Indeed if $c_1, c_2\in\mathbb{Q}$ such that $c_k=\gamma_{c_k}\cdot i\infty$, then $c_2=\gamma_{c_2}\gamma_{c_1}^{-1}\cdot c_1$. 
Next, let $c_2=m/n\in\mathbb{Q}$. It follows from the properties of the greatest common divisor that there are integers $p$ and $q$ such $\textrm{gcd}(m,n)=mp+nq$.\footnote{The integers $p,q$ can easily be obtained, e.g., from Mathematica's implementation {\tt GCD} of the greatest common divisor.} If $m$ and $n$ are relatively prime, $\textrm{gcd}(m,n)=1$, we can take $\gamma_{c_2}=\mat{m &mr-q}{n&nr+p}$, with $r$ an arbitrary integer. This matrix obviously satisfies $c_2=m/n=\gamma_{c_2}\cdot i\infty$. Moreover, this matrix has unit determinant, and so it defines an element of $\SL(2,\mathbb{Z})$.


\subsection{Modular forms}
\label{sec:modsub}

We start by considering functions from the extended upper half-plane into the complex numbers that are invariant under $\Gamma$. A \emph{modular function} for $\Gamma$ is a meromorphic function $f:{\overline{\mathbb{H}}}\to \mathbb{C}$ such that $f(\gamma\cdot \tau) = f(\tau)$, for all $\gamma\in\Gamma$. One can show that there are no non-constant holomorphic modular functions, i.e., every non-constant function that is invariant under $\Gamma$ has at least one pole. If we want to obtain holomorphic functions, we need to consider more general transformation behaviours under $\Gamma$. 
For every positive integer $n$, we define
\beq\label{eq:modular_action}
(f_{|n\gamma})(\tau) = (c\tau+d)^{-n}\,f(\gamma\cdot\tau)\,,\qquad \gamma =  \left(\begin{smallmatrix}a & b\\ c& d\end{smallmatrix}\right)\in\Gamma\,.
\eeq
This defines for every $n$ a genuine action of $\Gamma$ on functions on the extended upper half-plane. In particular, we have
\beq
(f_{|n\gamma_1})_{|n\gamma_2} = f_{|n(\gamma_1\gamma_2)}\,.
\eeq
A \emph{modular form of weight $n$ for $\Gamma$} is a function $f: \overline{\mathbb{H}}\to\mathbb{C}$ such that
\begin{enumerate}
\item $f$ invariant under the action of $\Gamma$ in eq.~\eqref{eq:modular_action}, i.e., it satisfies $(f_{|n\gamma})(\tau) =f(\tau)$.
\item $f$ is holomorphic on $\mathbb{H}$.
\item $f_{|n\gamma}$ is holomorphic at $i\infty$ for all $\gamma\in\SL(2,\mathbb{Z})$.
\end{enumerate}
We can think of modular forms as holomorphic functions, i.e., functions without poles, that are invariant under the action in eq.~\eqref{eq:modular_action}. Modular forms of weight $n$ form a finite-dimensional vector space $\cM_n(\Gamma)$. Moreover, $\cM_n(\Gamma)$ is a graded algebra, i.e., the product of two modular forms of weight $n_1$ and $n_2$ is a modular form of weight $n_1+n_2$. Since there are no non-constant holomorphic modular functions, all modular forms of weight zero are constant.

If $\Gamma$ is a congruence subgroup of level $N$ and $f\in\cM_n(\Gamma)$, then we say that $f$ has level $N$.  Every modular form of level $N$ is invariant under $T^N=\mat{1&N}{0&1}\in\Gamma(N)$. This matrix acts on $\overline{\mathbb{H}}$ via translations by $N$, $T^N\cdot \tau=\tau+N$, and so every modular form of level $N$  is a periodic function of period $N$ and admits a Fourier expansion of the form
\beq
f(\tau) = \sum_{n\ge 0} a_n\,q_N^n\,,\qquad q_N = e^{2\pi i\tau/N}\,.
\eeq

Having identified modular forms for $\Gamma$ as invariants under the group action in eq.~\eqref{eq:modular_action}, it is natural to ask how a modular form $f\in\cM_n(\Gamma)$ transforms for some $\gamma\in \SL(2,\mathbb{Z})$ that is not an element of $\Gamma$. 

We start by assuming that $\Gamma \subseteq \SL(2,\mathbb{Z})$ is a normal subgroup.\footnote{A subgroup $H\subseteq G$ is normal if $gH=Hg$, for all $g\in G$.} Then, for every $\gamma\in\SL(2,\mathbb{Z})$ and $\gamma_1\in \Gamma$, there is $\gamma_2\in\Gamma$ such that $\gamma\gamma_1=\gamma_2\gamma$. We have
\beq
(f_{|n\gamma})_{|n\gamma_1} = f_{|n(\gamma\gamma_1)} =  f_{|n(\gamma_2\gamma)} = (f_{|n\gamma_2})_{|n\gamma}  = f_{|n\gamma}\,,
\eeq
where the last step follows from the fact that $f$ is a modular form for $\Gamma$, and thus invariant. We conclude that if $f$ is a modular form for a normal subgroup $\Gamma$, then $f_{|n\gamma}$ is also a modular form for $\Gamma$ of the same weight, for all $\gamma\in \SL(2,\mathbb{Z})$.

The previous argument relies crucially on $\Gamma$ being normal in $\SL(2,\mathbb{Z})$. However, not all subgroups of $\SL(2,\mathbb{Z})$ are normal, not even all congruence subgroups. In particular, the congruence subgroups $\Gamma_0(N)$ and $\Gamma_1(N)$ are in general not normal. The principle congruence subgroups $\Gamma(N)$, however, are always normal. We can therefore generalise the previous statement in a weaker form to general congruence subgroups: If $f$ is a modular form for a congruence subgroup $\Gamma$ of level $N$, then $f_{|n\gamma}$ is a modular form for the principle congruence subgroup $\Gamma(N)$ of the same weight and level, for all $\gamma\in \SL(2,\mathbb{Z})$. We will see at the end of this section that it is in general not possible to make stronger statements, i.e., $f_{|n\gamma}$ will in general not be invariant under $\Gamma$, but it is only invariant under the smaller subgroup $\Gamma(N)$. 

While the previous argument shows that a modular form for a congruence subgroup $\Gamma$ of level $N$ in general transforms into a linear combination of modular forms for $\Gamma(N)$, it does not allow us to predict this linear combination. This situation changes if we restrict the discussion to a subset of modular forms, the so-called Eisenstein series, which we will study in more detail in the next section.


\subsection{Eisenstein series}
The vector space $\cM_n(\Gamma)$  of modular forms of weight $n$ for $\Gamma$ admits a direct sum decomposition
\beq
\cM_n(\Gamma) = \cE_n(\Gamma)\oplus \cS_n(\Gamma)\,.
\eeq
Here $\cS_n(\Gamma)$ denotes the subspace of cusp forms, i.e., the subspace of modular forms of weight $n$ for $\Gamma$ that vanish on $\mathbb{Q}\cup\{i\infty\}$. Its (orthogonal) complement is the Eisenstein subspace $\cE_n(\Gamma)$. So far mostly Eisenstein series have appeared in the context of Feynman integral calculations. We therefore study the structure of the Eisenstein subspaces in detail in the remainder of this section.

We first present an explicit basis for the space of Eisenstein series. There are numerous ways to write down a basis for $\cE_n(\Gamma)$ (cf.,~e.g.,~ref.~\cite{diamond}). Our choice is motivated by the fact that Eisenstein series for $\Gamma(N)$ are closely related to the Kronecker series and elliptic polylogarithms~\cite{BrownLevin,Broedel:2015hia,Broedel:2018iwv}. Since $\Gamma(N)\subseteq \Gamma$ for every congruence subgroup $\Gamma$ of level $N$, we only discuss the case of the principle congruence subgroup $\Gamma(N)$, and we refer to the literature for the other cases (see,~e.g.,~ref.~\cite{diamond}).

We start by writing down a spanning set for $\cE_n(\Gamma(N))$~\cite{Broedel:2018iwv},
\beq\label{eq:h_lattice_sum}
h_{N,r,s}^{(n)}(\tau) = -\sum_{\substack{(a,b)\in\mathbb{Z}^2\\ (a,b)\neq (0,0)}}\frac{e^{2\pi i(bs-ar)/N}}{(a\tau+b)^{n}}\,,
\eeq
where $n$ and $N$ are integers greater than unity and $r,s$ are integers defined modulo $N$. These functions transform like eq.~\eqref{eq:modular_action} and form a spanning set for $\cE_n(\Gamma(N))$, except for the case $n=2$ and $(r,s)=(0,0)\mod N$, which is not a modular form. More generally, we have~\cite{Broedel:2018iwv},
\beq\label{eq:h_trafo}
h_{N,r,s}^{(n)}\left(\frac{a\tau+b}{c\tau+d}\right) = (c\tau+d)^n\,h^{(n)}_{N,rd+sb,rc+sa}(\tau)\,,\qquad \gamma=\left(\begin{smallmatrix}a&b\\c& d\end{smallmatrix}\right)\in\SL(2,\mathbb{Z})\,.
\eeq

The functions defined in eq.~\eqref{eq:h_lattice_sum} admit a $q$-expansion that can be written down in closed form
\beq\label{eq:q-exp_closed}
h_{N,r,s}^{(n)}(\tau) = \sum_{k\ge 0}C_{N,r,s,k}^{(n)}\,q_N^k\,.
\eeq
The Fourier coefficients for $k\ge 1$ are given by (cf.,~e.g.,~ref.~\cite{diamond}),
\beq\label{eq:Fourier_h}
C_{N,r,s,k}^{(n)} = -\frac{(2\pi i)^{n}}{N^n(n-1)!}\sum_{(c_1,c_2)\in(\mathbb{Z}/N\mathbb{Z})^2}\sum_{\substack{m|k\\ k=mc_2}}\left[m^{n-1}\,e^{2\pi i(rc_2-(s-m)c_1)/N} - (m\leftrightarrow -m)\right]\,.
\eeq
The constant term for $n\ge2$ is
\beq\label{eq:Fourier_h_0}
C_{N,r,s,0}^{(n)} = \frac{(2\pi i)^n}{n!}\,B_n\left(\frac{s}{N}\right)\,,
\eeq
where $B_n(x)$ are the Bernoulli polynomials,
\beq
\frac{t\,e^{xt}}{e^t-1} = \sum_{n=0}^\infty\frac{t^n}{n!}\,B_n(x)\,.
\eeq
For $n=1$ the constant term is given by 
\beq
C_{N,r,s,0}^{(1)} = \left\{\begin{array}{ll}
2\pi i \left(\frac{s}{N}-\frac{1}{2}\right)\,, & s\neq 0\!\!\!\!\mod N\,,\\
0\,,& (r,s) = (0,0)\!\!\!\!\mod N\,,\\
\pi\cot\frac{\pi r}{N}\,,&\textrm{otherwise}\,.
\end{array}\right.
\eeq
Note that all the Fourier coefficients of $h_{N,r,s}^{(n)}(\tau)$ are algebraic multiples of $\pi^n$.

The functions $h_{N,r,s}^{(n)}$ form a spanning set of $\cE_n(\Gamma(N))$, but they are not linearly independent. It is known that for $n\ge 3$  the dimension of $\cE_n(\Gamma)$ is equal to the number of cusps of $\Gamma$.\footnote{For $n=2$, the dimension is equal to the number of cusps minus one, while for $n=1$ the dimension of $\cE_n(\Gamma(N))$ is only half the number of cusps.} The only linear relations among the $h_{N,r,s}^{(n)}$ for $n\ge 2$ are the reflection identity,
\beq\label{eq:reflection_id}
h_{N,r,s}^{(n)}(\tau) = (-1)^n\,h_{N,-r,-s}^{(n)}(\tau)\,,
\eeq
 and, for every $d$ divides $N$ and $0\le \rho,\sigma<N$, the distribution identity~\cite{Broedel:2018iwv},
 \beq\label{eq:distribution_id}
 \sum_{\frac{1}{N}(r,s)\in\frac{1}{N}(\rho,\sigma)+\Lambda_{N/d}^F}h_{N,r,s}^{(n)}(\tau) = \left(\frac{d}{N}\right)^{n-2}\,h^{(n)}_{d,\rho,\sigma}(\tau)\,,
 \eeq
 with 
 \beq
 \Lambda_{N/d}^F = \{(r,s)\in\mathbb{Z}^2:0\le r,s<N\}\,.
 \eeq
 These relations still hold for $n=1$, though in that case there are additional relations which can easily be found by looking for linear relations among the $q$-series. For $n\ge 3$, it is possible to explicitly write down a subset of the $h_{N,r,s}^{(n)}$ that are linearly independent~\cite{Broedel:2018iwv}.

Let us conclude this section by giving an example that if $f\in\cM_{n}(\Gamma)$ and $\gamma\in\SL(2,\mathbb{Z})$, then $f_{|n\gamma}$ will in general not be in $\cM_{n}(\Gamma)$, but only in $\cM_{n}(\Gamma(N))$.\footnote{We are grateful to Nils Matthes for suggesting this counterexample.} Consider the case $N=2$, and $\Gamma=\Gamma_1(2)$. The vector space $\cE_2(\Gamma(2))$ is two-dimensional, and a basis is $\{h^{(2)}_{2,1,0},h^{(2)}_{2,0,1}\}$. The subspace $\cE_2(\Gamma_1(2))\subset \cE_2(\Gamma(2))$ is one-dimensional with basis $\{h^{(2)}_{2,1,0}\}$. Consider $S=\left(\begin{smallmatrix}0&-1\\1&0\end{smallmatrix}\right)\notin \Gamma_1(2)$. We have
\beq
h^{(2)}_{2,1,0|6S}=h^{(2)}_{2,0,1}\,, \qquad h^{(2)}_{2,0,1|6S}=h^{(2)}_{2,1,0}\,,
\eeq
where the $h^{(2)}_{2,r,s|6S}$ refer to $h^{(2)}_{2,r,s}$ after acting with the modular transformation in eq.~\eqref{eq:modular_action} with $n=6$ and $\gamma=S$.
We see that the subspace $\cE_2(\Gamma_1(2))$ is not left invariant by $S$, because $h^{(2)}_{2,1,0|6S}\notin \cE_2(\Gamma_1(2))$.

\subsection{A parity-invariant spanning set}
\label{sec:parity_eis}
In the previous section we have presented a spanning set of Eisenstein series for $\Gamma(N)$ and we have described all the relations among these functions. This is important in applications, because it allows us to construct explicit bases for $\cE_n(\Gamma(N))$. The bases obtained in this way have the disadvantage that the resulting functions will in general be complex-valued. In applications, however, it can be useful to work with basis elements that are manifestly real, at least in certain regions of the parameter space. In this section we present a spanning set of functions for $\cE_n(\Gamma(N))$ that are explicitly real when $\tau$ is purely imaginary. 

We start by analysing how $h_{N,r,s}^{(n)}(\tau)$ behaves under complex conjugation when $\tau$ is purely imaginary. Writing $\tau=it$, with $t$ real and positive, we have
\beq\label{eq:h_parity}
h_{N,r,s}^{(n)}(it)^* = h_{N,r,-s}^{(n)}(it)\,.
\eeq
Indeed, starting from the lattice sum in eq.~\eqref{eq:h_lattice_sum}, we find
\beq
h_{N,r,s}^{(n)}(it)^*= -\sum_{\substack{(a,b)\in\mathbb{Z}^2\\ (a,b)\neq (0,0)}}\frac{e^{-2\pi i(bs-ar)/N}}{(-ait+b)^{n}}
= -\sum_{\substack{(\tilde{a},\tilde{b})\in\mathbb{Z}^2\\ (\tilde{a},\tilde{b})\neq (0,0)}}\frac{e^{-2\pi i(\tilde{b}s+\tilde{a}r)/N}}{(\tilde{a}it+\tilde{b})^{n}}
=h_{N,r,-s}^{(n)}(it)\,.
\eeq
We see that when $\tau$ is purely imaginary, complex conjugation amounts to changing the sign of $s$. Based on this observation, we define,
\beq\bsp\label{eq:ab_def}
\eisa^{(n)}_{N,r,s}(\tau) &\,= -\frac{1}{2}\left[h_{N,r,s}^{(n)}(\tau)+h_{N,r,-s}^{(n)}(\tau)\right]
=-\frac{1}{2}\left[h_{N,r,s}^{(n)}(\tau)+(-1)^nh_{N,-r,s}^{(n)}(\tau)\right]\,,\\
\eisb^{(n)}_{N,r,s}(\tau) &\,= -\frac{1}{2i}\left[h_{N,r,s}^{(n)}(\tau)-h_{N,r,-s}^{(n)}(\tau)\right]
=-\frac{1}{2i}\left[h_{N,r,s}^{(n)}(\tau)-(-1)^nh_{N,-r,s}^{(n)}(\tau)\right]\,,
\esp\eeq
where the second equality follows from the reflection identity in eq.~\eqref{eq:reflection_id}. 
These functions also admit a representation in terms of lattice sums similar to eq.~\eqref{eq:h_lattice_sum},
\beq\bsp
\eisa_{n,N,r,s}(\tau) &\,=\frac{1}{2}\sum_{\substack{(a,b)\in\mathbb{Z}^2\\ (a,b)\neq (0,0)}}\frac{e^{-2\pi iar/N}\cos\frac{2\pi sb}{N}}{(a\tau+b)^{n}}\,,\\
\eisb_{n,N,r,s}(\tau) &\,=\frac{1}{2i}\sum_{\substack{(a,b)\in\mathbb{Z}^2\\ (a,b)\neq (0,0)}}\frac{e^{-2\pi iar/N}\sin\frac{2\pi sb}{N}}{(a\tau+b)^{n}}\,.
\esp\eeq
Equation~\eqref{eq:h_parity} implies that the functions $\eisa_{n,N,r,s}(\tau)$ and $\eisb_{n,N,r,s}(\tau)$ are real whenever $\tau$ is purely imaginary. 
Since they are linear combinations of the Eisenstein series $h_{N,r,s}^{(n)}(\tau)$, they also form a spanning set for $\cE_n(\Gamma(N))$.We can easily determine a basis of $\eisa_{n,N,r,s}(\tau)$ and $\eisb_{n,N,r,s}(\tau)$. In particular, for $n\ge 3$ a basis for $\cE_n(\Gamma(N))$ is
\beq
B_{N,n} = B_{N,n}^1\cup B_{N,n}^2\,,
\eeq
with
\beq\bsp
B_{N,n}^1&\,=\left\{\eisa_{n,N,r,s},\eisb_{n,N,r,s}:0<r,s<N/2 \textrm{ and }\textrm{gcd}(N,r,s)=1\right\}\,.
\esp\eeq
The set $B_{N,n}^2$ depends on whether the weight $n$ is even or odd. If $n$ is even, we have
\beq
\!\!B_{N,n}^2=\left\{\eisa_{n,N,k,0},\eisa_{n,N,k,N/2},\eisa_{n,N,0,k},\eisa_{n,N,N/2,k}:0<k\le N/2 \textrm{ and }\textrm{gcd}(N,k)=1\right\}\,,
\eeq
while for $n$ odd, we have
\beq
\!\!B_{N,n}^2=\left\{\eisa_{n,N,k,0},\eisa_{n,N,k,N/2},\eisb_{n,N,0,k},\eisb_{n,N,N/2,k}:0<k\le N/2 \textrm{ and }\textrm{gcd}(N,k)=1\right\}\,.
\eeq
Note that when $N$ is odd, functions with index $N/2$ are absent.

%% file: ieis.tex

\section{Iterated integrals of modular forms}
\label{sec:ieis}

\subsection{Definition and basic properties}
\label{sec:ieis_props}

In this section we define our main objects of interest, namely iterated integrals of modular forms and iterated Eisenstein integrals. Consider a set of modular forms $h_j(\tau)$. We define differential forms $\omega_j = \frac{d\tau}{2\pi i}\,h_j(\tau)$ and consider the iterated integrals~\cite{ManinModular,Brown:mmv}, 
\beq\label{eq:IEI_to_II}
I(h_1,\ldots,h_k;\tau_0,\tau) = \int_{\tau_0}^{\tau}\omega_k\ldots\omega_1 = \int_{\tau_0}^{\tau}\frac{d\tau'}{2\pi i}\,h_1(\tau')\,I(h_2,\ldots,h_k;\tau_0,\tau') \,.
\eeq
The recursion starts with $I(;\tau_0,\tau)=1$.
The number of integrations $k$ is called the \emph{length} of the iterated integral. We can also define a notion of weight~\cite{Broedel:2018qkq}. Assume that all modular forms have level $N$ and that they admit $q$-expansions of the form
\beq
h_j(\tau) = \sum_{n\ge0}a_n(h_j)\,q_N^n\,,\qquad h_j \in\cM_{n_j}(\Gamma)\,,
\eeq 
where each Fourier coefficient $a_n(h_j)$ is a period of weight equal to $n_j$ (cf. for example the Fourier coefficients of the Eisenstein series in eq.~\eqref{eq:Fourier_h}, which are all algebraic multiples of powers of $\pi$). We define the weight of $I(h_1,\ldots,h_k;\tau_0,\tau)$ as $-k+\sum_{j=1}^kn_j$.

The integrals in eq.~\eqref{eq:IEI_to_II} have all the standard properties of iterated integrals, cf.~e.g., ref.~\cite{ChenSymbol}. In particular, they form a shuffle algebra,
\beq
I(h_1,\ldots,h_k;\tau_0,\tau)I(h_{k+1},\ldots,h_l;\tau_0,\tau) = \sum_{\sigma\in\Sigma(k,l-k)}I(h_{\sigma_1},\ldots,h_{\sigma_{l}};\tau_0,\tau)\,,
\eeq 
where the sum runs over all shuffles of $(h_1,\ldots,h_k)$ and $(h_{k+1},\ldots,h_l)$, i.e., over all permutations of their union that preserve the ordering within each set. The integrals satisfy the path composition and reversal formulas,
\beq\bsp\label{eq:path_comp}
I(h_1,\ldots,h_k;\tau_0,\tau) &\,= \sum_{l=0}^{k}I(h_{l+1},\ldots,h_k;\tau_0,\tau_1)I(h_{1},\ldots,h_l;\tau_1,\tau)\,,\\
I(h_1,\ldots,h_k;\tau_0,\tau) &\,= (-1)^kI(h_k,\ldots,h_1;\tau,\tau_0)\,.
\esp\eeq
It is easy to see that $I(h_1,\ldots,h_k;\tau_0,\tau)$ is linear in its first $k$ arguments,
\beq
I(\ldots,c_1\,h_{j,1} + c_2\,h_{j,2},\ldots;\tau_0,\tau) = c_1\,I(\ldots,h_{j,1},\ldots;\tau_0,\tau) + c_2\,I(\ldots,h_{j,2},\ldots;\tau_0,\tau)\,,
\eeq
where the $c_i$ are constants.
Finally, if all the modular forms $h_j$ are real on the imaginary axis, then $I(h_1,\ldots,h_k;\tau_0,\tau)$ is real whenever $\tau$ and $\tau_0$ are purely imaginary.

The path composition formula in eq.~\eqref{eq:path_comp} allows us to consider only a fixed base point $\tau_0$, and from now on we will always choose $\tau_0=i\infty$. We will use the shorthand notation $I(h_1,\ldots,h_k;\tau) = I(h_1,\ldots,h_k;i\infty,\tau)$. However, care is needed  to interpret these integrals, because strictly speaking they are divergent and require regularisation. In order to see that, we change variables from $\tau$ to $q_N$. We can write
\beq
\omega_j = \frac{dq_N}{q_N}\,\frac{N\,a_0(h_j)}{(2\pi i)^2} + \ord(q_N^0)\,. 
\eeq
Hence, if the constant term in the $q$-expansion is non zero, $\omega_j$ has a pole at $q_N=0$, or equivalently $\tau=i\infty$. As a result, the integral $I(h_1,\ldots,h_k;\tau)$ requires regularisation whenever $a_0(h_k)\neq 0$.

In ref.~\cite{Brown:mmv} it was shown how to replace logarithmically-divergent iterated integrals of modular forms by suitably regularised versions, such that all algebraic properties (e.g.,~the shuffle algebra) are preserved. This is achieved by interpreting eq.~\eqref{eq:IEI_to_II} as a linear combination of absolutely convergent integrals multiplied by powers of $\tau$~\cite{Brown:mmv},
\beq\label{eq:IEI_reg}
I(h_1,\ldots,h_k;\tau) = \sum_{p=0}^n\frac{a_{0}(h_{p+1})\ldots a_{0}(h_{k})}{(k-p)!\,(2\pi i)^{k-p}}\,\tau^{k-p}\int_{i\infty}^\tau R[\omega_p\ldots \omega_1]\,,
\eeq
where the map $R$ is defined by~\cite{Brown:mmv}
\beq
R[\omega_p\ldots \omega_1] = \sum_{q=0}^p(-1)^{p-q}\,\omega_{q+1}^{\infty}\ldots\omega_p^\infty\sh \omega_q\ldots\omega_1\,,
\eeq
where $\sh$ denotes the shuffle product, and we defined
\beq
\omega_j^{\infty}=a_0(h_j)\,\frac{d\tau}{2\pi i} =  \frac{dq_N}{q_N}\,\frac{N\,a_0(h_j)}{(2\pi i)^2}\,.
\eeq


\subsection{The elliptic ${}_2F_1$ function}
\label{sec:2F1_function}
In this section we illustrate the concepts from the previous section on a concrete example, namely the family of integrals defined by
\begin{align}\label{eq:2F1_def}
T(n_1,n_2,n_3;z) = \int_0^1 dx\, x^{-1/2 + n_1 + a\, \epsilon} (1-x)^{-1/2 + n_2 + b \epsilon} (1-z\, x)^{-1/2 + n_3 + c\,\epsilon}\,,
\end{align}
where $n_i\in\mathbb{Z}$ and $a,b,c$ are complex numbers. These integrals are closely related to a family of hypergeometric ${}_2F_1$ functions,
\beq\bsp
T(n_1,n_2,n_3;z) &\,= \frac{\Gamma(\frac{1}{2}+n_1+a\eps)\,\Gamma(\frac{1}{2}+n_2+b\eps)}{\Gamma(1+n_1+n_2+(a+b)\eps)}\\
&\,\times{}_2F_1\left(\frac{1}{2}-n_3-c\eps,\frac{1}{2}+n_1+a\eps;1+n_1+n_2+(a+b);z\right)\,.
\esp\eeq

The integral in eq.~\eqref{eq:2F1_def} defines a meromorphic function of $\eps$, and we interpret the integral as a Laurent series in $\eps$. In the following we assume that $z$ is real. As long as $z<1$ the integral is real order by order in $\eps$, but it develops an imaginary part when $z>1$. We assume that the branches of all functions are determined by assigning a small positive imaginary part to $z$. The dependence of the Laurent coefficients on $a$, $b$, $c$ is polynomial. For simplicity, we only discuss the case $a=b=c=1$. All conclusions remain true in the general case.

Using integration-by-parts identities, one can show that all integrals in the family in eq.~\eqref{eq:2F1_def} can be expressed as a linear combination of two independent master integrals, which can be chosen as
\begin{equation}\label{eq:2F1_masters}
T_1(z) = T(0,0,0;z) \textrm{~~and~~} \quad T_2(z) = T(1,0,0;z)\,.
\end{equation} 
The square root in the integrand defines an elliptic curve via the polynomial equation $y^2=x(1-x)(1-zx)$.
The master integrals were evaluated in ref.~\cite{Broedel:2017kkb,Broedel:2018iwv,Broedel:2018rwm} order-by-order in $\eps$ in terms of elliptic polylogarithms and iterated Eisenstein integrals for the congruence subgroup $\Gamma(2)$. 
The result can be cast in the form:
\beq\label{eq:T_S_U}
\left(\begin{array}{c}
T_1(z)\\ T_2(z)\end{array}\right) = S(z)
\left(\begin{array}{c}
U_1(\tau)\\ U_2(\tau)\end{array}\right)\,,
\eeq
where $S(z)$ denotes the matrix
\beq\label{eq:S_2F1_def}
S(z) =  \left(
\begin{array}{cc}
 2 \K(z) & 0 \\
 \frac{2}{z (1+6 \epsilon)}\,\left[\E(z)-\frac{1}{3} (2-z) \K(z)\right] & -\frac{i \pi }{z (1+6 \epsilon) \K(z)} \\
\end{array}
\right)\,.
\eeq
Here $\K(z)$ and $\E(z)$ denote the complete elliptic integrals of the first and second kind,
\beq
\K(z) = \int_0^1\frac{dx}{\sqrt{(1-x^2)(1-zx^2)}} \textrm{~~and~~} \E(z) = \int_0^1{dx}\,{\sqrt{\frac{1-zx^2}{1-x^2}}}\,.
\eeq
These integrals are real and well defined for $z<1$. Here we only focus on the region $0<z<1$ and we defer a discussion of the analytic continuation to the other regions to Section~\ref{sec:2F1}. In this region the functions $U_i(\tau)$ are power series in $\eps$ whose coefficients can be expressed in terms of iterated Eisenstein integrals. 
The variable $\tau\in\overline{\mathbb{H}}$ on which the iterated Eisenstein integrals depend is related to the variable $z$ in eq.~\eqref{eq:2F1_def} via
\beq\label{eq:tau_2F1_def}
\tau = i\,\frac{\K(1-z)}{\K(z)}\,.
\eeq
Note that $\tau$ is purely imaginary for $0<z<1$. 
The functions $\K(z)$ and $i\K(1-z)$ are in fact periods of the elliptic curve defined by the polynomial equation $y^2=x(1-x)(1-zx)$. The first few orders in the Laurent expansion of the functions $U_i(\tau)$ read
\beq\bsp
U_i(\tau) = \sum_{k=0}^\infty\eps^kU_{i,k}(\tau)\,,
\esp\eeq
with
\begin{align}
\nonumber U_{1,0}(\tau)&\,= 1\,,\\
\label{eq:U1i}U_{1,1}(\tau)&\,=8\, I(\eisa_{2,2,1,0};\tau)+4\, I(\eisa_{2,2,1,1};\tau)-2 \pi ^2 I(1;\tau )-4 \log2\,,\\
\nonumber U_{1,2}(\tau)&\,= \frac{\pi^2}{6} +8\log^22-16 \pi ^2\, I(1,\eisa_{2,2,1,0};\tau)-8 \pi ^2\, I(1,\eisa_{2,2,1,1};\tau)+180\, I(1,\eisa_{4,2,0,0};\tau)\\
\nonumber&\,-16 \pi ^2\, I(\eisa_{2,2,1,0},1;\tau)+64 \,I(\eisa_{2,2,1,0},\eisa_{2,2,1,0};\tau)+32\, I(\eisa_{2,2,1,0},\eisa_{2,2,1,1};\tau)\\
\nonumber&\,-8 \pi ^2\, I(\eisa_{2,2,1,1},1;\tau)+32\, I(\eisa_{2,2,1,1},\eisa_{2,2,1,0};\tau)+16\, I(\eisa_{2,2,1,1},\eisa_{2,2,1,1};\tau)\\
\nonumber&\,-32 \log2\, I(\eisa_{2,2,1,0};\tau)-16 \log2\, I(\eisa_{2,2,1,1};\tau)+8 \pi ^2 \log2\, I(1;\tau )\,,
\end{align}
and
\begin{align}
\nonumber U_{2,0}(\tau)&\,= 0\,,\\
\label{eq:U2i}U_{2,1}(\tau)&\,=i\pi\,,\\
\nonumber U_{2,2}(\tau)&\,=8 i \pi  I(\eisa_{2,2,1,0};\tau)+4 i \pi  I(\eisa_{2,2,1,1};\tau)+\frac{90}{i\pi }I(\eisa_{4,2,0,0};\tau)-4 i \pi  \log2\,.
\end{align}
Let us make some comments about these results. First, since $\tau$ is purely imaginary for $0<z<1$, we see that $U_1$ and $U_2$ are manifestly real and imaginary respectively (the explicit factor of $i$ in $U_2$ is cancelled by another explicit factor of $i$ in eq.~\eqref{eq:S_2F1_def}). Second, the functions $U_i$ are pure functions of uniform weight~\cite{ArkaniHamed:2010gh} in the sense of ref.~\cite{Broedel:2018qkq}. We emphasise that this statement depends on our choice for the periods of the elliptic curve: in eq.~\eqref{eq:S_2F1_def} we have singled out the periods $\K(z)$ and $i\K(1-z)$. We could of course have made a different choice of periods, and any two choices are related by an $\SL(2,\mathbb{Z})$ transformation,
\beq\label{eq:period_basis}
\left(\begin{array}{c}i\K(1-z)\\ \K(z)\end{array}\right) \longrightarrow \gamma \left(\begin{array}{c}i\K(1-z)\\ \K(z)\end{array}\right)\,,\qquad \gamma = \mat{a&b}{c&d} \in \SL(2,\mathbb{Z})\,,
\eeq
such that $\tau$ transform as in eq.~\eqref{eq:tau_trafo}. A different choice would have led to a different form for the matrix $S(z)$ and the pure functions $U_i(\tau)$. We will come back to this point in Section~\ref{sec:2F1} when we will discuss the analytic continuation of the integrals outside the range $0<z<1$. Here we only mention that, if we denote by $S(\gamma,z)$ the matrix corresponding to choosing another basis of periods (cf.~eq.~\eqref{eq:period_basis}), then this matrix will be related to $S(z)=S(\mathds{1},z)$ through the relation
\beq\label{eq:S_trafo}
S(\gamma,z) = S(z) \,M(\gamma^{-1},\tau)\,,
\eeq
with
\beq\label{eq:M_def}
M(\gamma,\tau) = \left(\begin{array}{cc}
c\tau+d&0\\
c & \frac{1}{c\tau+d}\end{array}\right)\,.
\eeq
Note that the matrices $M(\gamma,\tau)$ respect the group law of $\SL(2,\mathbb{Z})$,
\beq
M(\gamma_1\gamma_2,\tau) = M(\gamma_2,\tau)M(\gamma_1,\gamma_2\cdot\tau) \textrm{~~and~~} M(\gamma^{-1},\tau) = M(\gamma,\gamma^{-1}\cdot \tau)^{-1}\,.
\eeq

%% file: mpls.tex

\section{Expressing iterated Eisenstein integrals in terms of multiple polylogarithms}
\label{sec:eMPLs_to_MPLs}

In this section we discuss and clarify the relationship between iterated Eisenstein integrals and another class of iterated integrals that often show up in Feynman integral computations, namely multiple polylogarithms. We start by briefly reviewing multiple polylogarithms and their main properties, before we discuss their connection to iterated Eisenstein integrals in subsequent sections.

\subsection{Multiple polylogarithms}
Multiple polylogarithms (MPLs) are a generalisation of the ordinary logarithm and the classical polylogarithm function. They are defined by the iterated integrals~\cite{GoncharovMixedTate,Goncharov:2005sla},
\beq\label{eq:MPL_def}
G(a_1,\ldots,a_n;z) = \int_0^z\frac{dt}{t-a_1}\,G(a_2,\ldots,a_n;z)\,,
\eeq
where the $a_i$ and $z$ are (constant) complex numbers and the recursion starts from $G(;z)=1$. In the case $a_n=0$, the integral in eq.~\eqref{eq:MPL_def} is divergent, and we define instead,
\beq\label{eq:G_reg}
G(\underbrace{0,\ldots,0}_{n\textrm{ times}};z) = \frac{1}{n!}\log^nz\,.
\eeq
The number of integrations $n$ is called the weight of the MPL. MPLs contain the ordinary logarithm and the classical polylogarithms as special cases, e.g., 
\beq
G(\underbrace{0,\ldots,0}_{n-1\textrm{ times}},1;z) = -\textrm{Li}_n(z)\,.
\eeq

In order to understand the relation between MPLs and (certain) iterated Eisenstein integrals, it is instructive to understand the geometrical background leading to the definition of MPLs. Consider the space of rational functions in one variable $x$ with poles at most at some fixed positions $x=a_i \in \Sigma$ (we assume $\Sigma$ finite). It is easy to see that not every rational function of this type has a primitive within this space of functions, but we need to enlarge the space by including logarithms of the form $G(a_i,x)=\log(1-x/a_i)$. Indeed, we cannot find any rational function $R(x)$ such that its derivative equals $1/(x-a_i)$. This `obstruction' to finding a rational primitive is what naturally leads to the study of the (algebraic) de Rham cohomology groups $H^1_{\textrm{dR}}(X)$. In our case the space $X$ is the Riemann sphere with the points $a_i\in\Sigma$ removed, $X=\mathbb{CP}^1\setminus \Sigma$, and the `obstruction' to finding a rational primitive can be cast in the form of the statement that $H^1_{\textrm{dR}}(\mathbb{CP}^1\setminus \Sigma)$ is generated by the logarithmic one-forms $dx/(x-a_i)$,
\beq\label{eq:H1_CP1}
H^1_{\textrm{dR}}(\mathbb{CP}^1\setminus \Sigma) \simeq \left\langle \frac{dx}{x-a_i}, a_i\in\Sigma\right\rangle_{\mathbb{Q}}\,.
\eeq
The logarithmic one-forms are precisely the integration kernels that appear in the definition of the MPLs in eq.~\eqref{eq:MPL_def}.
Loosely speaking, the first de Rham cohomology group of a Riemann surface allows us to classify those meromorphic functions on $\mathbb{CP}^1\setminus \Sigma$ that we cannot trivially integrate in terms of meromorphic functions, but that require the introduction of new classes of functions. In the next section we will see that this cohomological intepretation allows us to express certain classes of iterated Eisenstein integrals in terms of MPLs.


\subsection{Iterated Eisenstein integrals and MPLs}

In Section~\ref{sec:modular_forms} we have defined the modular curve $Y_\Gamma=\mathbb{H}/\Gamma$. It can be shown that $Y_\Gamma$ is always a compact Riemann surface with a finite number of punctures, i.e., with a finite number of points removed. The punctures correspond to the cusps of $\Gamma$. A compact Riemann surface is characterised by its genus $g$, which, loosely speaking, counts the number of handles of the surface. Every compact Riemann surface of genus $g=0$ is isomorphic to the Riemann sphere $\mathbb{CP}^1$. In the previous section we have described the first de Rham cohomology group of a Riemann sphere with a finite number of points removed, and we have argued that the iterated integrals on that space are naturally connected to MPLs. It is therefore natural to expect that in cases where the modular curve $Y_{\Gamma}$ has genus zero, there is a connection between iterated integrals of modular forms and MPLs. In this section we briefly review this connection. The results of this section are well known in the mathematics literature. We review them here in a way appropriate for a physics audience.

Every meromorphic differential one-form on a Riemann surface, for example, every one-form of the form $dz\,R(z)$, where $z$ is a local coordinate on the Riemann surface and $R(z)$ is a meromorphic function, belongs to one of the following three classes:
\begin{itemize}
\item Differentials of the first kind are globally holomorphic, i.e., they have no poles anywhere on the Riemann surface.
\item Differentials of the second kind have poles with vanishing residues.
\item Differentials of the third kind have poles with non-vanishing residues.
\end{itemize}
The first de Rham cohomology group of a Riemann surface of genus $g$ is generated by precisely $g$ differentials of the first and $g$ differentials of the second kind, as well as one differential of the third kind for every puncture. In particular, for Riemann surfaces of genus zero the first cohomology group is entirely generated by differentials of the third kind, in agreement with eq.~\eqref{eq:H1_CP1}.

It turns out that there is a connection between modular forms of weight two for $\Gamma$ and the first de Rham cohomology group of $Y_\Gamma$. To see this, we start by noting that 
\beq
d(\gamma\cdot \tau) = \frac{d\tau}{(c\tau+d)^2}\,, \qquad \gamma=\mat{a&b}{c&d}\in\SL(2,\mathbb{Z})\,.
\eeq
In other words, $d\tau$ transforms like a modular form of weight $-2$. Hence, every modular form $h(\tau)$ of weight two defines a differential $d\tau\,h(\tau)$ that is invariant under $\Gamma$, and therefore furnishes a well-defined differential on $Y_{\Gamma}$. In addition, there is a one-to-one map between the differentials of the first kind on $Y_{\Gamma}$ (rather, the classes they define in the first de Rham cohomology group) and the cusp forms of weight two. In particular, this means that the genus of $Y_{\Gamma}$ is equal to the number of cusp forms of weight two for $\Gamma$. Similarly, there is a one-to-one map between the differentials of the third kind of $Y_{\Gamma}$ and modular forms of weight two from the Eisenstein subspace.

Assume now that we are given an iterated integral of Eisenstein series of weight two for some congruence subgroup $\Gamma$ for which $Y_{\Gamma}$ has genus zero, i.e., $Y_{\Gamma}$ is the Riemann sphere with a finite number of points removed. This happens precisely when $\textrm{dim }\cS_2(\Gamma) = 0$. For the congruence subgroups in eqs.~\eqref{eq:Gamma(N)_def} and~\eqref{eq:Gamma01(N)_def} it is well known for which level the modular curves have genus zero:
\begin{itemize}
\item $Y_0(N) = Y_{\Gamma_0(N)}$ has genus zero if and only if $N\in \{1,\ldots,10,12,13,16,18,25\}$.
\item $Y_1(N) = Y_{\Gamma_1(N)}$ has genus zero if and only if $N\in \{1,\ldots,10,12\}$.
\item $Y(N) = Y_{\Gamma(N)}$ has genus zero if and only if $N\in \{1,\ldots,5\}$.
\end{itemize}
In these cases there is a modular function $f:\mathbb{H}\to Y_{\Gamma}$ for $\Gamma$, called the \emph{Hauptmodul}, such that every modular function for $\Gamma$ is a rational function of $f(\tau)$. Hauptmodule are known for many of the modular curves considered above, see, e.g., refs.~\cite{Maier,Yang}. If $h(\tau)$ is an Eisenstein series of weight two for $\Gamma$, $\frac{d\tau}{2\pi i}\,h(\tau)$ is invariant under $\Gamma$, and $f$ maps it to a rational differential form on $Y_{\Gamma}$. Since we know all the generators of the first de Rham cohomology group of $Y_{\Gamma}$ from eq.~\eqref{eq:H1_CP1}, we can change variables to $t=f(\tau)$ and write $d\tau\,h(\tau)$ as a linear combination of logarithmic differential forms $dt/(t-t_i)$, where the $t_i=f(\tau_i)$ run over the images of the cusps of $\Gamma$ under the Hauptmodul $f$. The conclusion of this analysis can be summarised as follows:
\begin{quote}
If $\Gamma$ is a congruence subgroup such that $Y_{\Gamma}$ has genus zero, and $f:\mathbb{H}\to Y_{\Gamma}$ is a Hauptmodul for $\Gamma$, then every iterated Eisenstein integral of the form $I(h_1,\ldots,h_k;\tau)$ with $h_j\in\cE_2(\Gamma)$ can be written as a linear combination of uniform weight $k$ of MPLs evaluated at $t=f(\tau)$. 
\end{quote}

Let us illustrate this explicitly on the iterated Eisenstein integrals that appear in eqs.~\eqref{eq:U1i} and~\eqref{eq:U2i}. The relevant modular curve $Y(2)$ has genus zero, and a Hauptmodul is the modular lambda function,
\beq\label{eq:lambda_eta}
\lambda(\tau) = 16\,\frac{\eta(\tau/2)^8\eta(2\tau)^{16}}{\eta(\tau)^{24}}\,,
\eeq
where $\eta(\tau)$ denotes the Dedekind eta function,
\beq\label{eq:dedekind}
\eta(\tau) = q^{1/24}\,\prod_{n=1}^\infty(1-q^n)\,,\qquad q=e^{2\pi i\tau}\,.
\eeq
The modular lambda function is the inverse of eq.~\eqref{eq:tau_2F1_def} for $0<z<1$, i.e., $z=\lambda(\tau)$.
$\Gamma(2)$ has three cusps, which may be represented by $\tau=0$, $\tau=1$ and $\tau=i\infty$. The cusps are mapped by the Hauptmodul to the following points on the Riemann sphere,
\beq
\lambda(0) = 1\,,\qquad \lambda(1) = \infty\,,\qquad \lambda(i\infty) = 0\,.
\eeq
Hence, we can identify the modular curve $Y(2)$ with the Riemann sphere with the points $0$, $1$ and $\infty$ removed. We therefore expect that all iterated integrals of Eisenstein series of weight two and level two (except for $\eisa_{2,2,0,0}$) can be expressed in terms of MPLs in $z$ with singularities only when $z\in\{0,1,\infty\}$. The corresponding set of functions is a well-known subset of MPLs known as harmonic polylogarithms~\cite{Remiddi:1999ew}. In the remainder of this section we work this out in detail on the iterated integrals of Eisenstein series of weight two that appear in eqs.~\eqref{eq:U1i} and~\eqref{eq:U2i}.
Letting $z=\lambda(\tau)$, we can match $q$-expansions to show that
\beq\bsp\label{eq:tau_to_lambda}
\frac{d\tau}{2\pi i}\,\eisa_{2,2,1,0}(\tau) &\,= \frac{1}{12}\,\frac{dz}{z-1}-\frac{1}{6}\,\frac{dz}{z}\,,\\
\frac{d\tau}{2\pi i}\,\eisa_{2,2,1,1}(\tau) &\,= \frac{1}{12}\,\frac{dz}{z-1}+\frac{1}{12}\,\frac{dz}{z}\,.
\esp\eeq
Inserting these relations into the integrands of the iterated Eisenstein integrals and using the fact that the lower integration boundary $\tau=i\infty$ is mapped to $z=0$, we can easily perform all integrations in terms of MPLs.
We find
\beq\bsp
I(\eisa_{2,2,1,0};\tau) &\,= \frac{1}{12}G(1;z)-\frac{1}{6}G(0;z)+\frac{2}{3}\log2\,,\\
I(\eisa_{2,2,1,1};\tau) &\,= \frac{1}{12}G(1;z)+\frac{1}{12}G(0;z)-\frac{2}{3}\log2\,.
\esp\eeq
The constants proportional to $\log2$ require some explanation, because they cannot be recovered by inserting eq.~\eqref{eq:tau_to_lambda} into the integrand and performing the $z$ integration over the range $[0,z]$ in a naive way. In fact, this term has its origins in a different choice of regularisation scheme to regulate the logarithmic divergences at $\tau=i\infty$ and $z=0$, cf.~eqs.~\eqref{eq:IEI_reg} and~\eqref{eq:G_reg}. To understand this, we first note that if we expand eq.~\eqref{eq:lambda_eta} into a $q$-series, we find
\beq
z = \lambda(\tau) = 16\, q_2+\ord(q_2^2)\,,
\eeq
and so
\beq
G(0;z) = \log\lambda(\tau) = 4\,\log2+\log q_2 + \ord(q_2)\,.
\eeq
We see that terms proportional to (powers of) logarithms of $\log2$ arise from the fact that $z\sim 16\,q_2$ as $q_2\to 0$. 
These factors can always be recovered from the following recipe:
\begin{enumerate}
\item Insert eq.~\eqref{eq:tau_to_lambda} into the integrands of the iterated Eisenstein integrals and perform the integrals in the naive way.
\item Use the shuffle algebra properties to express every $G(a_1,\ldots,a_n;z)$ with $a_n=0$ in terms of MPLs with $a_n\neq0$ and powers of $G(0;z)$. 
\item Perform the replacement $G(0;z) \to G(0;z)-4\log2$. This replacement accounts for the difference in regularisation scheme.
\end{enumerate}
Using this procedure, we can cast the functions $U_i$ from Section~\ref{sec:2F1_function} in a simpler form involving only a handful of functions which cannot be expressed in terms of MPLs,
\beq\bsp\label{eq:U12_MPL}
U_1(\tau) &\,= 1+\epsilon\,  \left[G(1;z)-G(0;z)-2 \pi ^2 I(1;\tau)\right]\\
&\,+ \epsilon ^2\, \Bigg[180 \,I(1,\eisa_{4,2,0,0};\tau)+2 \pi ^2\left( G(0;z)-G(1;z)\right) I(1;\tau)\\
&\,+G(0,0;z)-G(0,1;z)-G(1,0;z)+G(1,1;z)+\frac{\pi ^2}{6}\Bigg]+\ord(\eps^3)\,,\\
U_2(\tau) &\,=i\pi\, \epsilon+i\pi\,\epsilon ^2\, \left[ G(1;z)-G(0;z)-\frac{90 }{\pi^2 }\,I(\eisa_{4,2,0,0};\tau)\right]+\ord(\eps^3)\,.
\esp\eeq

%% file: regularisation.tex

\section{Modular transformations of iterated integrals}
\label{sec:modular_trafo}

\subsection{Motivation}
At the end of the previous section we have given the first three orders in the $\eps$ expansion of the functions $U_1$ and $U_2$ which describe the familiy of integrals in eq.~\eqref{sec:2F1_function} in terms of MPLs and iterated Eisenstein integrals. The expressions in eq.~\eqref{eq:U12_MPL}, however, are only valid in the region where $0<z<1$. Indeed, in that region the variable $\tau$ defined in eq.~\eqref{eq:tau_2F1_def} is purely imaginary, because both $\K(z)$ and $\K(1-z)$ are real. If $z\notin[0,1]$, we need to analytically continue the elliptic integrals using the formula,
\beq
\K(z+i0) = \frac{1}{\sqrt{z}}\,\left[\K(1/z) + i\,\K(1-1/z)\right]\,,\quad z>1\,.
\eeq
For $z<0$, we obtain,
\beq
\tau = 1 + \frac{i\K\left(\frac{1}{1-z}\right)}{\K\left(\frac{z}{z-1}\right)} = \gamma_1\cdot\tau_1\,,
\eeq
with
\beq
\gamma_1=\mat{1&1}{0&1} \textrm{~~and~~} \tau_1=i\,\frac{\K\left(\frac{1}{1-z}\right)}{\K\left(\frac{z}{z-1}\right)}\,.
\eeq
Note that for $z<0$, $\tau_1$ is purely imaginary and $\textrm{Im}\,\tau_1>0$, and the matrix $\gamma_1$ lies in $\SL(2,\mathbb{Z})$. 
Similarly, for $z>1$, we find,
\beq\label{eq:tau3_def}
\tau = \frac{i\K\left(1-\frac{1}{z}\right)}{\K\left(\frac{1}{z}\right) + i\,\K\left(1-\frac{1}{z}\right)} = \gamma_3\cdot \tau_3\,,
\eeq
with
\beq
\gamma_3 = \mat{0&1}{-1&1} \textrm{~~and~~} \tau_3=i\frac{\K\left(\frac{1}{z}\right)}{\K\left(1-\frac{1}{z}\right)}\,.
\eeq
Again, we find that $\tau_3$ is purely imaginary with positive imaginary part for $z>1$, and that the matrix $\gamma_3$ lies in $\SL(2,\mathbb{Z})$.
Hence, if we define $\gamma_2=\mat{1&0}{0&1}$ and $\tau_2 = i\K(1-z)/\K(z)$ for $0<z<1$, we see that for each of the three regions there is $\gamma_i\in\SL(2,\mathbb{Z})$ and $\tau_i\in\mathbb{H}$ such that $\tau=\gamma_i\cdot \tau_i$. We can also define a map that assigns to all real values of $z$ the corresponding value of $\tau$,
\beq\label{eq:z_to_tau}
\tau(z) = \left\{\begin{array}{ll}
\gamma_1\cdot \tau_1\,,& \textrm{ if } z<0\,,\\
\gamma_2\cdot \tau_2\,,& \textrm{ if } 0\le z\le1\,,\\
\gamma_3\cdot \tau_3\,,& \textrm{ if } z>1\,.
\end{array}\right.
\eeq

In the remainder of this section we describe how we can express iterated Eisenstein integrals evaluated at $\tau=\gamma_i\cdot \tau_i$ in terms of iterated Eisenstein integrals in $\tau_i$. The motivations for doing this are twofold. First, when $z>1$, the integrals develop an imaginary part, and it can be desirable to make the imaginary parts explicit and only work with iterated integrals that are manifestly real. Since the Eisenstein series in eq.~\eqref{eq:ab_def} have the property of being real when evaluated at purely imaginary arguments, this will automatically be the case if we work with integrals evaluated at $\tau_i$. Second, as we will see in Section~\ref{sec:numerics}, we can speed up the numerical evaluation of the iterated Eisenstein integrals in different regions using modular transformation on its arguments.


\subsection{Modular transformations of iterated Eisenstein integrals}
The goal of this section is to describe a procedure to express iterated Eisenstein integrals of the form $I(h_1,\ldots,h_k;\gamma\cdot \tau)$, with $\gamma\in\SL(2,\mathbb{Z})$, in terms of iterated Eisenstein integrals evaluated at $\tau$.
The general strategy is simple: we start from the regularised version of the integral in eq.~\eqref{eq:IEI_reg}, perform the change of variables $\tau'\to\gamma\cdot\tau'$ in the integrand, and integrate back. Since $\SL(2,\mathbb{Z})$ is generated by the three elements $S$, $T$ and $-\mathds{1}$ in eq.~\eqref{eq:SL2Z_generators}, it is sufficient to discuss the cases $\gamma=S$ and $\gamma=T$, because a generic $\gamma\in\SL(2,\mathbb{Z})$ can always be written as a (finite) product of $S$'s and $T$'s, up to a sign.

Before we discuss the procedure in detail, we stress the importance of working with the regularised integrals in eq.~\eqref{eq:IEI_reg}: treating the divergent integrals naively can lead to wrong results!
Indeed, consider the integral
\beq\label{eq:I_G4_q}
I(\eisa_{4,1,0,0};\tau) = -\frac{\pi^2}{180}\,\log q_1 -\pi^2\,\left[\frac{4}{3}\,q_1+6q_1^2+\ord(q_1^3)\right]\,.
\eeq
Next, write $\tau=1+\tilde\tau=\mat{1&1}{0&1}\cdot \tilde\tau$. Using the definition of the iterated Eisenstein integrals, we can write,
\beq
I(\eisa_{4,1,0,0};1+\tilde\tau) = \int_{i\infty}^{1+\tilde\tau}\frac{d\tau'}{2\pi i}\,\eisa_{4,1,0,0}(\tau')\,.
\eeq
This integral diverges, and it needs to be interpreted as a regularised integral as discussed in Section~\ref{sec:ieis_props}. As we will illustrate now,  as a consequence of the regularisation, we cannot do naive changes of variables on the integral.
We can change variables according to $\tau'\to 1+\tau'$. The integrand is invariant under this change of variables, because $\eisa_{4,1,0,0}(1+\tau') = \eisa_{4,1,0,0}(\tau')$. Naively, the lower integration boundary $i\infty$ does not change under translations, and we are led to believe that $I(\eisa_{4,1,0,0};1+\tilde\tau)$ is identical to $I(\eisa_{4,1,0,0};\tilde\tau)$. It is easy to see from eq.~\eqref{eq:I_G4_q} that this cannot be correct: since $e^{2\pi i(\tilde\tau+1)}=e^{2\pi i\tilde\tau} = q_1$, we see that the power series part in eq.~\eqref{eq:I_G4_q} remains unchanged. The first term, however, does change, and we find,
\beq\label{eq:reg_example}
I(\eisa_{4,1,0,0};1+\tilde\tau) = I(\eisa_{4,1,0,0};\tilde\tau) -\frac{i\pi^3}{90}\,.
\eeq
We see that the constant in eq.~\eqref{eq:reg_example} is a consequence of the regularisation.
Indeed, if we start from the regularised version in eq.~\eqref{eq:IEI_reg}, we find
\beq
I(\eisa_{4,1,0,0};\tau) = -\frac{i\pi^3}{90}\,\tau + \int_{i\infty}^\tau\frac{d\tau'}{2\pi i}\,\left[\eisa_{4,1,0,0}(\tau')-\frac{\pi^4}{45}\right]\,.
\eeq
The second term is an absolutely convergent integral and gives rise to the power series part in eq.~\eqref{eq:I_G4_q}. We can naively perform the change of variables and conclude that it is invariant under translations. The first term immediately produces a constant offset under translations, and so we recover eq.~\eqref{eq:reg_example}.

We can apply exactly the same steps to more general modular transformations: we always start from the regularised version in eq.~\eqref{eq:IEI_reg} and perform a change of variables in the absolutely convergent integrals. 
Before we proceed, however, we need to restrict the class of iterated integrals that we want to consider.
First, we note that we can always use the shuffle algebra to transform all iterated integrals into a new basis of integrals where none of the $\omega_j$ correspond to a modular form of weight zero (except for explicit powers of $\tau$), but instead we consider differential forms of the form $d\tau\,h_j(\tau)\,\tau^{m_j}$, for some positive integer $m_j$ and $h_j$ a modular form of weigh $n_j$. An example will clarify this:
\beq\bsp
I(1,h,1;\tau) &\,= I(1;\tau)\,I(h,1;\tau)-2I(h,1,1;\tau)\\
&\,=\frac{\tau}{2\pi i}\int_{i\infty}^\tau \frac{d\tau'}{2\pi i}\,h(\tau')\,I(1;\tau')-2\int_{i\infty}^\tau \frac{d\tau'}{2\pi i}\,h(\tau')\,I(1,1;\tau')\\
&\,=\frac{\tau}{2\pi i}\int_{i\infty}^\tau \frac{d\tau'}{2\pi i}\,h(\tau')\,\frac{\tau'}{2\pi i}-\int_{i\infty}^\tau \frac{d\tau'}{2\pi i}h(\tau')\,\left(\frac{\tau'}{2\pi i}\right)^2\,.
\esp\eeq
From here on, and for the rest of this paper, we will restrict ourselves to iterated integrals where, after transforming the integrals to this new basis, only differential forms $d\tau\,h_j(\tau)\,\tau^{m_j}$ with $0\le m_j<n_j-1$ and $n_j>1$ appear. The reason for this restriction will become clear shortly. The same restriction has already appeared in ref.~\cite{ManinModular}. We emphasise that this restriction is not a problem for the applications that we have in mind, because in all known physics applications only iterated Eisenstein integrals for this restricted class appear.

Let us now consider an integral from this restricted class, and let us work out the change of variables explicitly. We write down formulas valid for arbitrary $\gamma\in\SL(2,\mathbb{Z})$, 
though we keep in mind that we can recover every modular transformation via a suitable sequence of $S$ and $T$ transformations.
If $\gamma = \left(\begin{smallmatrix}a& b\\c& d\end{smallmatrix}\right)\in\SL(2,\mathbb{Z})$, we find,
\beq\bsp\label{eq:example_I}
I(h_1,\ldots,h_k;\gamma\cdot\tau) &\,= \sum_{p=0}^n\frac{a_{0}(h_{p+1})\ldots a_{0}(h_{k})}{(2\pi i)^{k-p}(k-p)!}\,(\gamma\cdot \tau)^{k-p}\int_{i\infty}^{\gamma\cdot\tau} R[\omega_p\ldots \omega_1]\\
&\,= \sum_{p=0}^n\frac{a_{0}(h_{p+1})\ldots a_{0}(h_{k})}{(2\pi i)^{k-p}(k-p)!}\,\left(\frac{a\tau+b}{c\tau+d}\right)^{k-p}\int_{-d/c}^{\tau} R[\omega_p^\gamma\ldots \omega_1^\gamma]\,,
\esp\eeq
where $\omega_j^{\gamma}$ denotes the transform of $\omega_j$ under $\gamma$,
\beq
\omega_j =\frac{d\tau}{2\pi i}\,h_j(\tau) \xrightarrow{\,\,\,\gamma\,\,\,}\omega_j^{\gamma} =\frac{d\tau}{2\pi i}\,{(c\tau+d)^{-2}}\,h_j(\gamma\cdot \tau)\,.
\eeq
In the previous section we have seen that if $h_j(\tau)$ is a modular form of weight $n_j$ for $\Gamma(N)$, then so is $h_j(\gamma\cdot \tau)$. In particular, if $h_j = h_{N,r,s}^{(n_j)}$, eq.~\eqref{eq:h_trafo} implies,
\beq\label{eq:example_h}
\omega_j^\gamma = \frac{d\tau}{2\pi i}\,(c\tau+d)^{n_j-2}\,h^{(n_j)}_{N,rd+sb,rc+sa}(\tau)\,.
\eeq
The transformation behaviour of the Eisenstein series $\eisa_{n,N,r,s}$ and $\eisb_{n,N,r,s}$ can easily be recovered from their definition in terms of $h_{N,r,s}^{(n)}$ and $h_{N,r,-s}^{(n)}$, cf.~eq.~\eqref{eq:ab_def}. 
The additional powers of $\tau$ that are generated in the right-hand side can be expressed in terms of iterated integrals of modular forms of weight 0, 
\beq
\tau^p=(2\pi i)^p\,p!\,I(\underbrace{1,\ldots,1}_{p};\tau)\,.
\eeq
We can then simply integrate back in terms of (regulated) iterated integrals of modular forms to obtain the desired result. 
We stress that at this point it is crucial that we work with the restricted class of integrals: indeed, if we change to the basis of integrals where the differential forms have the form $d\tau\,h_j(\tau)\,\tau^{m_j}$, then these differential forms transform as
\beq\label{eq:condition}
d\tau\,h_j(\tau)\,\tau^{m_j} \xrightarrow{\,\,\,\gamma\,\,\,} d\tau\,h_j(\tau)\,(a\tau+b)^{m_j}\,(c\tau+d)^{n_j-m_j-2}\,.
\eeq
We see that no negative powers appear precisely when the condition $0\le m_j\le n_j-2$ is fulfilled. This condition also automatically enforces $n_j\ge 2$, which are precisely the conditions that define our restricted class of integrals. We note that this condition is preserved under modular transformations~\cite{ManinModular} (which can be seen, for example, from the fact that the highest power of $\tau$ in eq.~\eqref{eq:condition} never exceeds $n_j-2$ for $0\le m_j\le n_j-2$). In other words, we have described an algorithm that allows us to derive modular transformations of iterated integrals of modular forms from this restricted class, and this restricted class is closed under modular transformations.


Second, we see from eq.~\eqref{eq:example_I} that the lower integration boundaries are constant rational values $-\frac{d}{c}=\gamma^{-1}\cdot i\infty$. If we split the path of integration so that all lower integration limits are the infinite cusp, then we are naturally led to consider iterated Eisenstein integrals evaluated at constant rational values. If we decompose $\gamma$ into a product of $S$'s and $T$'s, we only obtain constant iterated Eisenstein integrals evaluated at $\tau=0$, because
\beq
T\cdot i\infty = i\infty {\textrm{~~and~~}} S\cdot i\infty = 0\,.
\eeq
Iterated Eisenstein integrals evaluated at $\tau=0$ are related to a class of transcendental numbers that has recently appeared in the mathematics literature. We will review this class of transcendental constants in the remainder of this section.

For modular forms $h_j$, $1\le j\le k$, of weight $n_j\ge 2$ and integers $m_j$ such that $1\le m_j<n_j$, the associated \emph{multiple modular value} (MMV) is defined by the integral~\cite{ManinModular,Brown:mmv}:
\beq\bsp\label{eq:mmv_def}
\Lambda&(h_k,\ldots,h_1;m_k,\ldots,m_1)=\\
&\, = (-i)^{m}\,\int_0^{i\infty}d\tau_1\,h_1(\tau_1)\,\tau_1^{m_1-1}\int_{0}^{\tau_1}\ldots\int_{0}^{\tau_{k-1}}d\tau_k\,h_k(\tau_k)\,\tau_k^{m_k-1}\,,
\esp\eeq
with $m=m_1+\ldots+m_k$. We note that the restrictions on $m_j$ and $n_j$ that appear in the definition of MMVs are precisely the conditions that define our restricted class of integrals.
The integral in eq.~\eqref{eq:mmv_def} may diverge, and we have to interpret it again as a regularised version. We define the weight of an MMV as $n_1+\ldots+n_k$ and the length as $m$. 
MMVs of small weight and lengths associated to modular forms of level one have been studied in ref.~\cite{Brown:mmv,Brown:mmv2}. It is easy to see that MMVs form a shuffle algebra,
\beq\bsp\label{eq:Lambda_def}
\Lambda&(h_l,\ldots,h_{k+1};m_l,\ldots,m_{k+1})\,\Lambda(h_k,\ldots,h_1;m_k,\ldots,m_1) \\
&\,= \sum_{\sigma\in\Sigma(k,l)}\Lambda(h_{\sigma_l},\ldots,h_{\sigma_1};m_{\sigma_k},\ldots,m_{\sigma_1})\,,
\esp\eeq
where the sum runs over all shuffles of $k$ and $l-k$ elements. Let us conclude by discussing an identity relating different MMVs. The $S$-transformation maps the integration path $[i\infty,0]$ to itself (up to reversal of the orientation), and so this transformation induces a transformation on MMVs.
We also note that there is a reflection identity for MMVs associated to Eisenstein series for $\Gamma(N)$,
\beq
\Lambda(h_{N,r_k,s_k}^{(n_k)},\ldots,h_{N,r_1,s_1}^{(n_1)};m_k,\ldots,m_1) = i^n\,\Lambda(h_{N,\tilde{r}_1,\tilde{s}_1}^{(n_1)},\ldots,h_{N,\tilde{r}_k,\tilde{s}_k}^{(n_k)};\tilde{m}_1,\ldots,\tilde{m}_k)\,,
\eeq
where $n$ is the weight of the MMV, and 
\beq
\tilde{m}_j=n_j-m_j\,,\qquad \tilde{r}_j = N-s_j\,,\qquad \tilde{s}_j = r_j\,.
\eeq
The reflection identity can easily be obtained by working out the transformation of the corresponding iterated integrals under the modular transformation $S:\tau\to -1/\tau$.

MMVs are (conjecturally transcendental) constants. In applications it is important to obtain their numerical value. Since MMVs are special cases of iterated integrals of modular forms, namely those evaluated at $\tau=0$, we can try to proceed in the same way as for iterated integrals and evaluate them numerically using $q$-expansions. In this case, however, the expansion parameter is $q_N=1$, which leads to badly converging expansions. In the next section we describe a way to accelerate the convergence of the series.

%% file: numerics.tex

\section{Asymptotic expansions and numerical evaluation of iterated Eisenstein integrals}
\label{sec:numerics}

In this section we discuss how to efficiently evaluate Eisenstein series and their iterated integrals. A key ingredient is the behaviour of iterated Eisenstein integrals under modular transformations discussed in Section~\ref{sec:modular_trafo}. Our strategy is to derive fast converging $q$-series close to every cusp. We therefore start by discussing how to obtain asymptotic expansions of Eisenstein series and their iterated integrals in general, before we turn to the numerical evaluation in subsequent sections.

\subsection{Asymptotic expansions of iterated Eisenstein integrals}
\label{sec:asymp_exp}
In applications it is often useful to expand Feynman integrals into an asymptotic series, e.g., close to a singular point of the differential equation satisfied by the integral.  When transformed to the variable $\tau$, the singular points of the (homogeneous) differential equation correspond to the cusps of the modular curve. We are therefore particularly interested in asymptotic expansions close to a cusp. 

We start by discussing $q$-expansions of Eisenstein series, and we comment on iterated integrals at the end of the section.
We already know that every modular form admits a $q$-expansion around the infinite cusp at $\tau=i\infty$ (or $q_N=0$). For Eisenstein series we can write down the $q$-expansion around the infinite cusp in closed form, cf.~eq.~\eqref{eq:q-exp_closed}. We now discuss how we can obtain the $q$-expansion around other cusps. We note that this is well-known in the mathematics literature, but we feel that it is important to discuss this topic here for completeness. 

Consider $c\in\mathbb{Q}$ and an Eisenstein series $f\in \cE_n(\Gamma(N))$. Since the $h^{(n)}_{N,r,s}(\tau)$ form a spanning set for $\cE_n(\Gamma(N))$, it is sufficient to consider the case where $f$ is an element of this set.\footnote{We could also work with the spanning set formed by the $\eisa_{n,N,r,s}$ and $\eisb_{n,N,r,s}$. The conversion between the two sets is trivial, cf.~eq.~\eqref{eq:ab_def}.} We have seen in Section~\ref{sec:modular_forms} that there is $\gamma_c\in\SL(2,\mathbb{Z})$ such that $c=\gamma_c\cdot i\infty$. Hence, expanding $h^{(n)}_{N,r,s}(\tau)$ around the cusp $\tau=c$ is equivalent to expanding $h^{(n)}_{N,r,s}(\gamma_c\cdot\tau')$ around $\tau'=i\infty$ using eqs.~\eqref{eq:h_trafo} and~\eqref{eq:q-exp_closed}. Let us illustrate this on an example: Assume we want to expand $f(\tau) = h^{(4)}_{2,1,1}(\tau)$ into a series close to $\tau=1/2$. We have $\gamma_{1/2} = \mat{1&0}{2&1}$, and eq.~\eqref{eq:h_trafo} gives
\beq\label{eq:num_ex_1}
h^{(4)}_{2,1,1}(\tau) = h^{(4)}_{2,1,1}(\gamma_{1/2}\cdot\tau') = (2\tau'+1)^4\,h^{(4)}_{2,1,1}(\tau')\,.
\eeq
Using eq.~\eqref{eq:q-exp_closed} we can obtain the desired expansion,
\begin{align}\label{eq:num_ex_2}
h^{(4)}_{2,1,1}(\tau)  &\,= (2\tau'+1)^4\,\frac{\pi^4}{360}\left[7 + 240\,e^{i\pi\tau'}-240\,e^{2i\pi\tau'}+6720\,e^{3i\pi\tau'}+\ldots\right]\\
\nonumber&\,=(2\tau-1)^{-4}\,\frac{\pi^4}{360}\left[7 + 240\,e^{i\pi\tau/(1-2\tau)}-240\,e^{2i\pi\tau/(1-2\tau)}+6720\,e^{3i\pi\tau/(1-2\tau)}+\ldots\right].
\end{align}

Using this approach, we can obtain asymptotic expansions close to every cusp (e.g., close to every singular point of the differential equation satisfied by the Feynman integral). The strategy for obtaining asymptotic expansions of iterated Eisenstein integrals follows exactly the same lines: we can work out the modular transformation properties of the integrals using the steps described in Section~\ref{sec:modular_trafo} to map any $c\in\mathbb{Q}$ to the infinite cusp, where we can easily obtain asymptotic expansions. Fast converging expansions are also the basis for the numerical evaluation of Eisenstein series and their iterated integrals. We turn to this topic in the next section.


\subsection{Numerical evaluation}
\label{sec:num_eval}

In this section we discuss how to evaluate Eisenstein series and their iterated integrals numerically in an efficient way. We focus only on Eisenstein series, but all statements can easily be transposed to iterated Eisenstein integrals. 

Consider an Eisenstein series $h_{N,r,s}^{(n)}(\tau)$. This function admits a $q$-expansion, and we can easily write down an arbitrary number of terms in the expansion, cf.~eq.~\eqref{eq:q-exp_closed}. Whenever $\textrm{Im}\,\tau>0$, we have $|q_N|<1$, and the series is convergent. However, the convergence may be rather slow depending on the value of $q_N$. For example, if $\tau\sim i\infty$ (and $\textrm{Re}\,\tau$ is not too large), we have $|q_N|\ll1$ and the series converges extremely fast. However, when $\textrm{Im}\,\tau$ is not large, the convergence of the series can be extremely slow. We illustrate this in the second column of Table~\ref{tab:h4211_num}, where we show the numerical value obtained for $h_{2,1,1}^{(4)}(\tau)$ at $\tau=\frac{1}{2}+\frac{i}{10}$ after truncating the $q$-expansion around the infinite cusp after a certain number of terms. Te expansion parameter is $e^{i\pi \tau}\simeq i\,0.7304\ldots$.

\begin{table}[!t]
\begin{center}
\begin{tabular}{c|c|c}
\hline\hline
 $n_{\textrm{max}}$ & direct & transformed \\
 \hline
 1   & $36.538 + 47.432i$ &  $1183.791 - 15.756i$ \\
 2   & $36.538 + 47.432i$ &  $1183.797 - 15.756i$ \\
 3   & $36.538 - 661.090i$ & $1183.797 - 15.756i$\\ 
 4   & $18.056 - 661.090i$ & $1183.797 - 15.756i$\\
 5   & $18.056 + 1039.856i$ &  $ 1183.797 - 15.756i$\\
 10 & $642.471 + 1470.891i$ &  $   1183.797 - 15.756i$\\
 30 &  $1172.467 + 51.758i$&    $ 1183.797 - 15.756i$\\
 60 & $1183.787 - 15.799i$ & $ 1183.797 - 15.756i$\\
 70 & $1183.797 - 15.753i$&  $1183.797 - 15.756i$\\
 75 & $1183.797 - 15.757i$ & $1183.797 - 15.756i$\\
 \hline\hline
 \end{tabular}
 \caption{\label{tab:h4211_num}Numerical evaluation of $h_{2,1,1}^{(4)}(\tau)$ at $\tau=\frac{1}{2}+\frac{i}{10}$ using a truncated $q$-expansion with $n_{\textrm{max}}$ terms. The column labeled `direct' is obtained by expanding $h_{2,1,1}^{(4)}(\tau)$ around $\tau=i\infty$, while the numbers in the column labeled `transformed' are obtained by mapping $\tau$ to the fundamental domain before expansion.} 
  \end{center}
 \end{table}
 
 The convergence of the series can be considerably accelerated by using modular transformations. For every $\tau\in\mathbb{H}$, there is $\gamma\in\SL(2,\mathbb{Z})$ such that $\gamma\cdot \tau$ lies in the fundamental domain for $\SL(2,\mathbb{Z})$ (cf.~eq.~\eqref{eq:fundamental_domain}). In Appendix~\ref{app:SL2Z} we review how this matrix can be constructed. In the case of our example, we find
\beq
\frac{1}{2}+\frac{i}{10} = \gamma_{1/2}\cdot\left(-\frac{1}{2}+\frac{5i}{2}\right)\,,
\eeq
where the matrix $\gamma_{1/2}=\mat{1&0}{2&1}$ of Section~\ref{sec:modular_trafo}. Using the transformation in eq.~\eqref{eq:num_ex_1} and the expansion in eq.~\eqref{eq:num_ex_2}, we obtain a series representation with expansion parameter $e^{i\pi\tau/(1-2\tau)}\simeq -i\,0.000388\ldots$. The convergence of this series is extremely fast, as can be seen from the third column in Table~\ref{tab:h4211_num}.
 
Using this strategy, we can derive fast converging series representations for Eisenstein series and their iterated integrals for all points in the upper half-plane. There is, however, one potential issue which we need to address. Our strategy involves using modular transformations in order to map a point in $\mathbb{H}$ to the fundamental domain before expansion. In Section~\ref{sec:modular_trafo} we have argued that modular transformations introduce new transcendental constants, and we have argued that these constants are MMVs. Thus, if our strategy for efficient numerical evaluation is supposed to work, we need to be able to obtain numerical approximations for the new transcendental constants. Since all the MMVs relevant here are iterated Eisenstein integrals evaluated at $\tau=0$, we can use the same strategy to obtain numerical values for these MMVs. At this point, however, we need to address a critical point: since the iterated integrals are evaluated at $\tau=0$, a naive $q$-expansion would have $q_N=1$ as expansion parameter, and the $q$-series converges very badly. We can use modular transformations to accelerate the convergence, but this procedure may introduce new MMVs that we need to evaluate. We thus seem to be stuck in an infinite loop. In the following we discuss how we can circumvent this problem and evaluate all MMVs we need without introducing any new transcendental constants. 

The basic idea is very simple. We can split the path of integration $[0,i\infty]$ into the segments $[0,i]$ and $[i,i\infty]$. The path $[0,i]$ can be mapped to $[i\infty,i]$ using $S$, because $S\cdot0= i\infty$ and $S\cdot i=i$.
Hence, using the path composition and reversal formul\ae, as well as the algorithm from the previous section, we can express every MMV as a linear combination of iterated integrals of modular forms evaluated at $\tau=i$. The expansion parameter for the $q$-expansion is then $q_N = e^{-2\pi/N}<1$, and we obtain a fast-converging series representation of the MMV. After these steps all paths extend from $i\infty$ to $i$, and we do not introduce any new transcendental constants. We can use this procedure to evaluate all MMVs we need to high precision. We have checked that we reproduce all results for MMVs of level one of small weight and length of ref.~\cite{Brown:mmv2}. The precision we can reach is only limited by the number of terms in the expansion one wishes to include.  We have also computed all MMVs up to weight six associated to Eisenstein series for $\Gamma(2)$, as well as all MMVs associated to Eisenstein series for $\Gamma(6)$ through weight five. In all cases, we were able to obtain at least 100 digits. Using the PSLQ algorithm, we find that all of these MMVs can be expressed in terms of the transcendental constants related to MPLs evaluated at the sixth root of unity, except for two and three new constants of weight four and five respectively.

%% file: F21.tex

\section{Application to the elliptic ${}_2F_1$ function}
\label{sec:2F1}

In the remainder of this paper we illustrate the concepts of the previous sections on several examples. We start in this section by returning to the elliptic ${}_2F_1$ function of Section~\ref{sec:2F1_function}, before we discuss examples of Feynman integrals that involve iterated Eisenstein integrals in the next section. 

\subsection{The imaginary part of the elliptic ${}_2F_1$ function}

So far we have only considered the integrals in eq.~\eqref{eq:2F1_def} for $0<z<1$, where the integral is real. We have already mentioned that for $z>1$ the integral develops an imaginary part  (we assume that all branches are fixed by assigning a small positive imaginary part to $z$). In this section we show how we can use modular transformations of iterated Eisenstein integrals to make this imaginary part explicit.

The idea we will follow is very simple. We have seen in eq.~\eqref{eq:tau3_def} that for $z>1$ we can write $\tau=\gamma_3\cdot \tau_3$, with $\gamma_3\in\SL(2,\mathbb{Z})$ and $\tau_3$ purely imaginary. We can then write all integrals in terms of iterated Eisenstein integrals evaluated at $\tau_3$. If we work with the parity-invariant spanning set of Eisenstein series of Section~\ref{sec:parity_eis}, all iterated Eisenstein integral are real for $z>1$. Hence, we can explicitly isolate the imaginary parts of the integrals. 

In the following we only discuss the functions $U_i(\tau)$ defined in eq.~\eqref{eq:U12_MPL}. The integrals $T_1$ and $T_2$ can be recovered by multiplying with the matrix $S(z)$ in eq.~\eqref{eq:S_2F1_def}, which does not involve any iterated Eisenstein integrals. For the functions $U_i(\tau)$, we find
\beq\label{eq:2F1_imag}
\left(\begin{array}{c}
U_1(\tau) \\ U_2(\tau)
\end{array}\right) = \left(\begin{array}{c}
U_1(\gamma_3\cdot \tau_3) \\ U_2(\gamma_3\cdot \tau_3)
\end{array}\right)
=M(\gamma_3,\tau_3)^{-1}\,\left(\begin{array}{c}\tilde U_1(\tau_3)+i\pi\, \tilde  V_{1}(\tau_3)\\ \tilde U_2(\tau_3)+i\pi\, \tilde  V_2(\tau_3)
\end{array}\right)\,,
\eeq
where the matrix $M(\gamma_3^{-1},\tau_3)$ was defined in eq.~\eqref{eq:M_def}. The real parts are given by
\begin{align}
\nonumber\tilde{U}_1(\tau_3) &\,=1+\epsilon\,\Bigg[G(1;{1}/{z})-2\pi^2\,I(1;\tau_{3})\Bigg]+\epsilon^2\,\Bigg[180I(1,\eisa_{4,2,0,0};\tau_{3})\\
&\,-2\pi^2G(1;{1}/{z})\,I(1;\tau_{3})+G(1,1;{1}/{z})-\frac{\pi^2}{3}\Bigg]+\ord(\eps^3)\,,\\
\nonumber\tilde{U}_2(\tau_3)&\,=-1-\epsilon\,G(1;{1}/{z})+\epsilon^2\,\Bigg[\frac{5\pi^2}{3}-180I(\eisa_{4,2,0,0},1;\tau_{3})-G\left(1,1;{1}/{z}\right)\Bigg]+\ord(\eps^3)\,.
\end{align}
The imaginary parts are
\begin{align}
\nonumber\tilde{V}_1(\tau_3) &\,=-2I(1;\tau_{3})-2\epsilon\,G(1;{1}/{z})\,I(1;\tau_{3})-2\epsilon^2\Bigg[180\,I(1,\eisa_{4,2,0,0},1;\tau_{3})\\
&\,+G(1,1;{1}/{z})\,I(1;\tau_{3})+\frac{15}{\pi^2}\zeta_3\Bigg]+\ord(\eps^3)\,,\\
\nonumber\tilde{V}_2(\tau_3) &\,=\epsilon + \epsilon^2\Bigg[G(1;{1}/{z})-\frac{90}{\pi^2}\,I(\eisa_{4,2,0,0};\tau_{3})\Bigg]+\ord(\eps^3)\,.
\end{align}
Since $\tau_{3}$ is purely imaginary for $z>1$, the functions $\tilde{U}_i$ and $\tilde{V}_i$ are manifestly real in this region. Therefore this is the correct separation of $U_i(\tau)$ into its real and imaginary parts.

Let us make some comments about these results. First we note that the functions are still pure functions of uniform weight, i.e., the property of uniform weight is preserved. The functions $U_i$, however, are no longer pure after analytic continuation: indeed, the matrix $M(\gamma_3,\tau_3)$ in eq.~\eqref{eq:2F1_imag} contains $\tau_{3}$ in the denominator, which spoils purity. However, all the terms that spoil purity are contained in the matrix $M(\gamma_3,\tau_3)$. This can be explained as follows: The separation into the matrix $S(z)$ and the pure functions $U_i(\tau)$ in eq.~\eqref{eq:T_S_U} relies on a choice, namely the choice of two periods for the elliptic curve $y^2=x(1-x)(1-zx)$ associated to our problem, or equivalently the choice of $\tau$ in eq.~\eqref{eq:tau_2F1_def}. Of course, we could have chosen a different basis of periods, and any two choices are related by an $\SL(2,\mathbb{Z})$ transformation. The matrices $S(z)$ obtained from two different choices are related by the matrix $M(\gamma_3,\tau_3)$, cf.~eq.~\eqref{eq:S_trafo}. Since nothing should depend on this choice, the matrix $M(\gamma_3,\tau_3)$ in eq.~\eqref{eq:S_trafo} must be cancelled by a similar contribution coming from the modular transformation of the iterated Eisenstein integrals, so that 
\beq\label{eq:T_S_U}
\left(\begin{array}{c}
T_1(z)\\ T_2(z)\end{array}\right) = S(z)
\left(\begin{array}{c}
U_1(\tau)\\ U_2(\tau)\end{array}\right)
= S(\gamma_3,z)
\left(\begin{array}{c}\tilde U_1(\tau_3)+i\pi\, \tilde  V_{1}(\tau_3)\\ \tilde U_2(\tau_3)+i\pi\, \tilde  V_2(\tau_3)
\end{array}\right)\,.
\eeq
This explains the appearance of the matrix $M(\gamma_3,\tau_3)$ in eq.~\eqref{eq:2F1_imag}.


\subsection{Numerical evaluation of the elliptic ${}_2F_1$ function}

In the previous section we have seen how to analytically continue the master integrals for the family of elliptic ${}_2F_1$ functions to all real values of $z$. After analytic continuation all imaginary parts are explicit, but the resulting expressions like in eq.~\eqref{eq:2F1_imag} are not necessarily in a form where we can evaluate all the functions in a fast and stable way. Indeed, we have seen in Section~\ref{sec:num_eval} that when $\tau$ has a small imaginary part, the $q$-expansions converge very slowly.

\begin{figure}[!t]
\begin{center}
\includegraphics[scale=0.2]{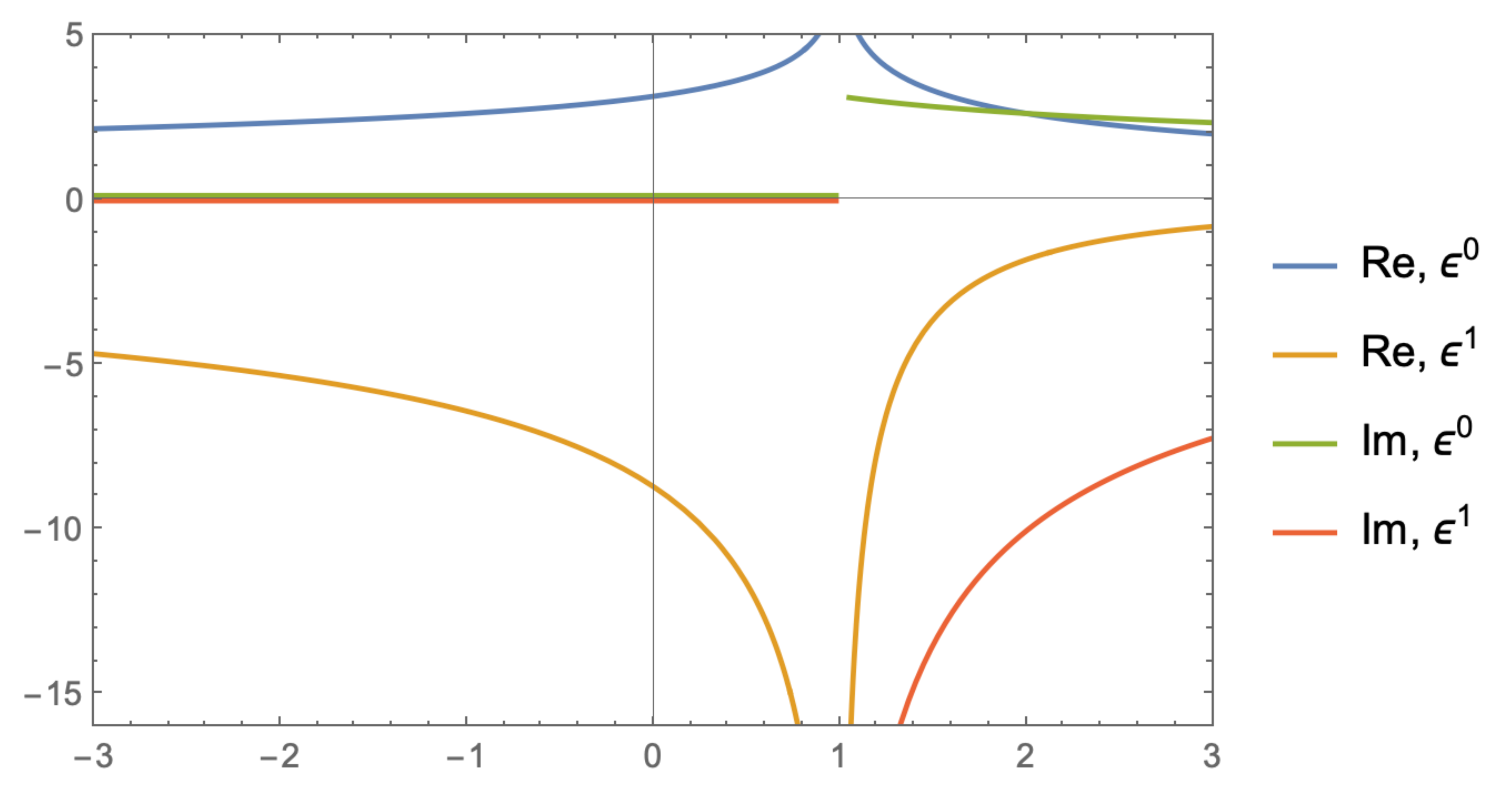}
\caption{\label{fig:T1}The real and imaginary parts of the first three orders in the $\eps$ expansion of the master integral $T_1$.}
\end{center}
\end{figure}

Alternatively, for every (real) value of $z$, we can find $\gamma_z\in\SL(2,\mathbb{Z})$ such that $\gamma_z^{-1}\cdot \tau(z)$ lies in the fundamental domain, where $\tau(z)$ was defined in eq.~\eqref{eq:z_to_tau}.
A priori, the matrix $\gamma_z$ depends on $z$. We need only six distinct values of $\gamma_z$ to cover all real values of $z$:
\beq\label{eq:gamma_z_def}
\gamma_z=\left\{\begin{array}{ll}
\gamma_{1a} \,,& \textrm{ if }z\le -1\,,\\
\gamma_{1b} \,,& \textrm{ if }-1<z<0 \,,\\
\gamma_{2a} \,,& \textrm{ if }0\le z<1/2 \,,\\
\gamma_{2b} \,,& \textrm{ if }1/2\le z\le 1 \,,\\
\gamma_{3a} \,,& \textrm{ if }1< z\le2 \,,\\
\gamma_{3b} \,,& \textrm{ if }z>2 \,,
\end{array}\right.
\eeq
where we defined
\beq\bsp
\gamma_{1a} &\,=\mat{-1&1}{-1&0}\,, \qquad\,\,\, \phantom{\gamma_1}\gamma_{2a} =\gamma_2=\mat{1&0}{0&1}\,,  \qquad\,\,\,\, \gamma_{3a} =\mat{0&1}{-1&1}\,,\\
\gamma_{1b} &\,=\gamma_1=\mat{1&1}{1&0}\,, \qquad \gamma_{2b} =\mat{0&-1}{1&0}\,,  \qquad \gamma_{3b} =\gamma_3=\mat{-1&0}{-1&-1}\,.
\esp\eeq
It is easy to check that the variables $\tau_{j\alpha}\equiv \gamma_{j\alpha}\cdot \tau(z)$, $j\in\{1,2,3\}$ and $\alpha\in\{a,b\}$, are purely imaginary and $\textrm{Im }\tau_{j\alpha}\ge i$. In each of the six regions in eq.~\eqref{eq:gamma_z_def} we obtain a representation of the master integrals $T_i$ in eq.~\eqref{eq:2F1_masters} in terms of iterated Eisenstein integrals evaluated at $\tau_{j\alpha}$, which admit fast converging $q$-expansions. The corresponding representations of $T_1$ and $T_2$ are shown in Appendix~\ref{app:2f1_results}. As an illustration of the numerical evaluation, we show in Fig.~\ref{fig:T1} a plot of the real and imaginary parts of the first three coefficients in the Laurent expansion of $T_1$.

Let us conclude this section by commenting on the six regions for $z$ that appear in eq.~\eqref{eq:gamma_z_def}. The values $\{0,1\}$ may not come as a surprise, because they are related to the singular points of the integral. The values $\{-1,1/2,2\}$ need more explanation, because they do not seem to be directly related to any apparent special point of the integral. In order to understand the appearance of these points, we need to analyse the $j$-invariant of the elliptic curve $y^2=x(1-x)(1-zx)$ associated to our problem. The periods and $\tau$ contain redundant information, but the $j$-invariant is a complex number that uniquely characterises an elliptic curve. The $j$-invariant can be computed from $\tau$. It is invariant under $\SL(2,\mathbb{Z})$ (i.e., it is a modular function for $\SL(2,\mathbb{Z})$), and it can be expressed as a rational function of Eisenstein series of level one:
\beq
j(\tau) = \frac{20\,h_{1,0,0}^{(4)}(\tau)^3}{20\,h_{1,0,0}^{(4)}(\tau)^3+49\,h_{1,0,0}^{(6)}(\tau)^2}\,.
\eeq
Since $j(\tau)$ is a modular function for $\SL(2,\mathbb{Z})$, it must be a rational function of the Hauptmodul $z=\lambda(\tau)$. We have
\beq
j(\tau) = \frac{4 \left(1-z+z^2\right)^3}{27\, (1-z)^2 \,z^2}\,.
\eeq
We now see that the poles of the $j$-invariant are $z\in\{0,1\}$, while $z\in\{-1,1/2,2\}$ corresponds to $j(\tau)=1$! The corresponding values of $\tau$ are
\beq
\lambda(1+i) = -1\,,\qquad \lambda(i) = 1/2\,,\qquad  \lambda\left(\frac{1}{2}+\frac{i}{2}\right)=2\,.
\eeq
We quote this here merely as an observation, we will see a similar pattern in the next section in the case of elliptic Feynman integrals, where the different regions will be determined by the points where the $j$-invariant becomes infinite, zero or one.


\subsection{Asymptotic expansions}
Let us conclude this section with an illustration how we can obtain asymptotic expansions of the elliptic ${}_2F_1$ functions order by order in $\eps$. In Section~\ref{sec:asymp_exp} we have shown how we can expand Eisenstein series and their iterated integrals into a $q$-series close to a cusp. In many applications, however, one would like to have expansions in terms of the original variables of the problem (e.g., kinematic variables in the case of Feynman integrals). 
In this section we illustrate how we can obtain expansions in $z$ close to the singular points $z=0$ or $z=1$.

In the following we only discuss in detail the expansion of $T_1(z)$ and $T_2(z)$ close to $z=0$. The case $z=1$ can be dealt with in the the same way. We have $\tau(0) = i\infty$, and so expanding for small values of $z$ is equivalent to expanding around $\tau(z)\sim i\infty$. We can easily expand all iterated Eisenstein integrals in eq.~\eqref{eq:U12_MPL} into a $q$-series around $\tau=i\infty$ and then convert the $q$-series into a series in $z$ by expanding eq.~\eqref{eq:tau_2F1_def}. We find
\beq
\tau(z) = \frac{i}{\pi }\,(4 \log 2-\log z)-\frac{i}{2 \pi }\,z-\frac{13 i }{64 \pi }\,z^2-\frac{23 i}{192 \pi }\,z^3 -\frac{2701 i}{32768 \pi }\,z^4+\ord(z^5)\,.
\eeq
or equivalently,
\beq\label{eq:q2_to_z}
q_2 = e^{i\pi\tau(z)} = \frac{z}{16}+\frac{z^2}{32}+\frac{21\, z^3}{1024}+\frac{31\, z^4}{2048}+\ord(z^5)\,.
\eeq
Inserting eq.~\eqref{eq:q2_to_z} into the $q$-series, we find
\beq\bsp
T_1(z)&\, = \pi\,\left[1+\frac{1}{4}\,z+\frac{9}{64}\,z^2+\ord(z^3)\right]\\
&\,-\epsilon\,\pi\,  \Bigg[4  \log2+z \left(\frac{1}{2}+  \log2\right)+z^2 \left(\frac{27  }{64}+\frac{9}{16}   \log2\right)+\ord(z^3)\Bigg]\\
&\,+\epsilon^2 \,\pi\,\Bigg[\frac{\pi ^2}{6}+8  \log ^22+z \left(\frac{\pi ^2}{24}+2  \log ^22+2   \log2\right)\\
&\,\qquad\qquad+z^2 \left(\frac{23  }{64}+\frac{3 \pi ^2}{128}+\frac{9}{8}   \log ^22+\frac{27}{16}   \log2\right)+\ord(z^3)\Bigg]+\ord(\eps^3)\,,\\
T_2(z)&\,=\pi\,\Bigg[\frac{1 }{3 z}-\frac{1 }{12}-\frac{11 }{192}\,z-\frac{29 }{768}\,z^2+\ord(z^3)\Bigg]\\
&\,+\epsilon \,\pi\, \Bigg[-\frac{4}{3 z}\,\log2-\frac{1 }{6}+\frac{1}{3}   \log2
+z \left(\frac{25  }{192}+\frac{11}{48}   \log2\right)\\
&\,\qquad\qquad+z^2 \left(\frac{283  }{2304}+\frac{29}{192}   \log2\right)+\ord(z^3)\Bigg]\\
&\,+\epsilon ^2 \,\pi\,\Bigg[\frac{1}{z}\,\left(\frac{\pi ^2}{18}+\frac{8}{3}   \log ^22\right)
-\frac{\pi ^2}{72}-\frac{2}{3}   \log ^22+\frac{2}{3}   \log2\\
&\,\qquad\qquad+z \left(-\frac{13  }{192}-\frac{11 \pi ^2}{1152}-\frac{11}{24}   \log ^22-\frac{25}{48}   \log2\right)\\
&\,\qquad\qquad+z^2 \left(-\frac{35  }{256}-\frac{29 \pi ^2}{4608}-\frac{29}{96}   \log ^22-\frac{283}{576}   \log2\right)
\Bigg]+\ord(\eps^3)\,.
\esp\eeq
This expansion is the same as the one obtained from the usual series representation of Gauss' hypergeometric function,
\beq
{}_2F_1(a,b;c;z) = \sum_{n=0}^\infty\frac{(a)_n(b)_n}{(c)_n}\,\frac{z^n}{n!}\,.
\eeq

%% file: sunrise.tex

\section{Application to the sunrise integral}
\label{sec:sunrise}

As a further application of the ideas described in the previous sections, we would like to perform the analytic continuation
for the class of Feynman integrals required to compute the sunrise graph.
In order to keep the formulas as compact as possible, we will limit ourselves to consider the relevant master integrals
in $d=2$ dimensions, where the master integrals can be chosen to be finite. We recall that the value in $d=4$ dimensions can be obtained by a straightforward
application of the well-known formulas for the dimensional shift of Feynman integrals~\cite{Tarasov:1996br,Lee:2009dh}.

We will start with a short recap on the computation of the sunrise graph, which will also allow us to introduce our notation. 
We consider the family of two-loop integrals defined by the formula
\beq\bsp
\label{eq:sunrise-fam}
S_{a_1,\dots, a_5}&(p^2, m^2; d) \\
&\,=   \int \ddl_1\ddl_2 \frac{(\ell_1\cdot p)^{a_4} (\ell_2\cdot p)^{a_5}}{[\ell_1^2 - m^2]^{a_1} [\ell_2^2 - m^2]^{a_2}  [(\ell_1-\ell_2-p)^2-m^2]^{a_3}}\ ,
\esp\eeq
where the integration measure is chosen as
\begin{equation}
  \label{eqn:intemeasure}
  \int\ddl=\frac{1}{\Gamma\left( 2-\frac{d}{2} \right)}\int\frac{d^d\ell}{i\pi^{d/2}}\,.
\end{equation}
Using IBP identities~\cite{Chetyrkin:1981qh,Tkachov:1981wb}, one can show that the family of integrals defined in eq.~\eqref{eq:sunrise-fam} can be reduced to three master integrals.
We choose one of the master integrals to be the tadpole, which is one in our normalisation,
\begin{equation}
S_{2,2,0,0,0}(p^2, m^2; 2-2\eps) = 1 \,.
\end{equation}
Following ref.~\cite{Remiddi:2016gno,Broedel:2019kmn}, we choose the other two master integrals as follows
\beq\bsp
\label{eq:sunrise-basis}
\cS_1 (\eps; t) &= -m^2\, S_{1,1,1,0,0}(p^2, m^2; 2-2\eps) \ ,\\
\cS_2 (\eps; t) &= -\left[ \frac{1}{3}(t^2-6t+21) -12\eps(t-1)\right]
    m^2\, S_{1,1,1,0,0}(p^2, m^2; 2-2\eps)\\
    &\quad- 2 (t-1)(t-9) m^4\, S_{2,1,1,0,0}(p^2, m^2; 2-2\eps),
\esp\eeq
where we defined the dimensionless variable $t = p^2 /m^2$ and rescaled by the appropriate powers of $m$
in order to render $\cS_1 (\eps; t)$ and $\cS_2 (\eps; t)$ dimensionless as $\eps \to 0$. 
Note that both integrals are finite in $d=2$ dimensions.
From now on, for simplicity, we will set $m=1$, as the dependence of the master integrals on $m$ can be always be recovered
from simple dimensional analysis.

As it is well known, the two non-trivial master integrals of the two-loop massive sunrise graph satisfy a system of two coupled
differential equations. With this choice of basis, the equations in $d=2$ space-time dimensions read~\cite{Remiddi:2016gno}
\beq
\bsp
\label{eq:sunrise-deq}
\partial_t \begin{pmatrix} \cS_1 ( t)  \\ \cS_2 (t) \end{pmatrix} \,=\, B(t)  \begin{pmatrix} \cS_1 (t)   \\ \cS_2 (t)  \end{pmatrix}  + \begin{pmatrix}0 \\ 1 \end{pmatrix}\ ,
\esp
\eeq
with
\beq\bsp
\label{eq:sunrise-deq-matrices}
B(t)\,&=\, \frac{1}{6 \,t\, (t-1) (t-9)}
\begin{pmatrix}
3(3+14t-t^2) & -9 \\
(t+3)(3+75 t -15 t^2 + t^3)  & -3(3+14t -t^2)
\end{pmatrix}\,,
\esp\eeq
and where we defined $\cS_j ( t) = \cS_j (0; t)$ with $j=1,2$. Equation~\eqref{eq:sunrise-deq} is a coupled $2\times 2$ system of equations whose
solution requires one  to first find a full solution of the corresponding homogeneous system, 
i.e., a $2 \times 2$ matrix $\mathcal{W}_S(t)$ such that
\beq
\partial_t \mathcal{W}_S(t) = B(t) \mathcal{W}_S(t)\,.
\eeq
As shown for the first time in ref.~\cite{Laporta:2004rb}, and generalised in ref.~\cite{Primo:2016ebd}, 
a particular solution can always be found by studying the maximal cut of the relevant graphs.
In our case, we can use the first-order differential operator 
\beq
\mathcal{D}_t = \tfrac{1}{3}(3+14t-t^2)-\tfrac{2}{3}(t-9)(t-1)\partial_t\,,
\eeq
to write $\mathcal{W}_s(t)$ in the region $0<t<1$ as
\begin{equation}\label{eq:sunrise_wronskian}
    \mathcal{W}_S(t) = \begin{pmatrix}
        \Psi_1(t) & \Psi_2(t) \\
        \mathcal{D}_t\Psi_1(t) & \mathcal{D}_t\Psi_1(t)\\
    \end{pmatrix}\,,
\end{equation}
where
\beq\bsp
  \label{eq:psi1_def}
  \Psi_1(t) & = \frac{4}{[(3-\sqrt{t})(1+\sqrt{t})^3]^{1/2}}\,{\K}\left(\frac{t_{14}(t)t_{23}(t)}{t_{13}(t)t_{24}(t)}\right)\,,\\
  \Psi_2(t) & = \frac{4 i}{[(3-\sqrt{t})(1+\sqrt{t})^3]^{1/2}}\,{\K}\left(\frac{t_{12}(t)t_{34}(t)}{t_{13}(t)t_{24}(t)}\right)\,,
\esp\eeq
are the two periods of the associated elliptic curve, with $t_{ij}(t) = t_i(t)-t_j(t)$ and the four branching points are
\beq\label{eq:SR_t_i_def}
t_1(t) = -4\,,\quad t_2(t) = -(1+\sqrt{t})^2\,,\quad t_3(t) = -(1-\sqrt{t})^2\,, \quad t_4(t)=0\,.
\eeq
At this point, we can define a new set of master integrals $\cT_1(t),\; \cT_2(t)$ as follows
\begin{equation} 
\begin{pmatrix} \cS_1(t) \\ \cS_2(t) \end{pmatrix} =
\cW_S(t) \begin{pmatrix}\cT_1(t)\\ \cT_2(t) \end{pmatrix}\,, \label{eq:SinT}
\end{equation}
which by definition fulfil the purely non-homogeneous system of differential equations
\beq\label{eq:sunrise_T_2}
\partial_t\left(\begin{matrix}\cT_1(t)\\ \cT_2(t)\end{matrix}\right) =\frac{1}{4 \pi i} \begin{pmatrix} -\Psi_2(t) \\ \phantom{-}\Psi_1(t) \end{pmatrix}\,.
\eeq

It is by now well known that a solution of these differential equations can be found in terms of iterated
integrals of Eisenstein series for $\Gamma_1(6)$~\cite{Adams:2017ejb,Broedel:2019kmn}. The modular parameter $\tau$ 
is as usual defined as the ratio of the two periods,
\begin{equation}
  \label{eq:psi3def}
  \tau(t)=\frac{\Psi_2(t)}{\Psi_1(t)}\,,
\end{equation}
and we find that the $j$-invariant of the family of elliptic curves associated to the maximal cut of the sunrise graph reads
\begin{equation}
\label{eq:jinv_sun}
j(t) = \frac{\left(t^4-12 t^3+30 t^2+228 t+9\right)^3}{1728 (t-9)^2 (t-1)^6 t}\,.
\end{equation}
It  turns out that eq.~\eqref{eq:psi3def} can be inverted to give
\begin{equation}
  \label{eq:t16}
  t(\tau)=9\frac{\eta(\tau)^4\eta(6\tau)^8}{\eta(2\tau)^8\eta(3\tau)^4}\,,
\end{equation}
where $\eta(\tau)$ was defined in eq.~\eqref{eq:dedekind}.
The function $t(\tau)$ is a Hauptmodul for $\Gamma_1(6)$, and so it is invariant under the corresponding modular transformations.
$\Psi_1(t)$, instead, is a modular form of weight $n=1$ for $\Gamma_1(6)$. Starting from these observations, one can prove
that a basis for $\mathcal{M}_n(\Gamma_1(6))$ is given by the functions~\cite{Broedel:2018rwm}
\beq\label{eq:Gamma_1(6)_basis}
f_{n,p}(\tau) = \Psi_1(t(\tau))^n\,t(\tau)^p\,,\quad 0\le p\le n\,,
\eeq
where we assume
$f_{0,0}(\tau)=1$. 
Finally we introduce the following notation for iterated integrals of modular forms for $\Gamma_1(6)$,
\beq
\IMFo{n_1 &\ldots & n_k \\ p_1&&p_k}{\tau} = I(f_{n_1,p_1},\ldots,f_{n_k,p_k};\tau) = \int_{i\infty}^{\tau}d\tau'\,f_{n_1,p_1}(\tau')\,\mathcal{I}(f_{n_2,p_2},\ldots,f_{n_k,p_k};\tau')\,.
\eeq
With these definitions it is immediate to see that as long as $0<t<1$, 
the sunrise graph in $d=2$ dimensions can be written in terms of iterated integrals of modular forms for $\Gamma_1(6)$:
\beq\bsp
\label{eq:finalresult-sunrise}
\mathcal{T}_1(t(\tau)) &\,= \frac{{\rm Cl}_2(\pi/3)}{2\pi} -\frac{1}{24\pi^2 }\,\left[\IMFo{3 & 0\\ 3&0}{\tau} - 10\, \IMFo{3 & 0\\ 2&0}{\tau} + 9\,\IMFo{3 & 0\\ 1&0}{\tau}\right]\,,\\
\mathcal{T}_2(t(\tau)) &\,= \frac{1}{24\pi^2}\,\left[\IMFo{3 \\ 3}{\tau} - 10\, \IMFo{3 \\ 2}{\tau} + 9\,\IMFo{3\\ 1}{\tau}\right]\,,
\esp\eeq
where
\beq
{\rm Cl}_2(x)\,=\, \frac{i}{2}(\text{Li}_2(e^{-i x})-\text{Li}_2(e^{i x}))\ .
\eeq
is the Clausen function. We stress that eq.~\eqref{eq:finalresult-sunrise} is only valid in the region $0<t<1$, which corresponds to $0<s<m^2$.
For these values of the external invariants the two master integrals of the sunrise graph 
$\mathcal{S}_1$, $\mathcal{S}_2$ are real, which remains true as long as $s < 9 m^2$ (including
negative values of $s$).  On the other hand, the integrals have a branch cut starting at $s=9m^2$, and they develop an imaginary part for $s > 9 m^2$. Note that this is in general not true for $\mathcal{T}_1$ and $\mathcal{T}_2$ separately.

Before we discuss how to rewrite this solution for different values of $s$, we will express it 
in terms of iterated integrals over  the  parity-invariant spanning set of modular forms for $\Gamma(6)$ 
defined in Section~\ref{sec:parity_eis}. 
By now it should be clear that this is the natural starting point to be able to analytically continue our expression to the whole phase space. 
Our goal will be to perform modular transformations 
in order to remap $\tau$ to the fundamental domain, whenever needed. Once that is done, we can evaluate the relevant
iterated integrals by very fast converging $q$-expansions. 
Equation~\eqref{eq:finalresult-sunrise} is written in terms of modular forms for
 $\Gamma_1(6)$, which is not a normal subgroup of $\SL(2,\mathbb{Z})$ and, as it was discussed in Section~\ref{sec:modsub},
 we do not expect this set of iterated integrals to be closed under  $\SL(2,\mathbb{Z})$ transformations. $\Gamma(6)$, instead,
 is a normal subgroup of $\SL(2,\mathbb{Z})$ and it is therefore the natural arena where we expect our solution to be defined 
 for general values of $s$.
 Moreover, by using the parity-invariant spanning set we are also able to make all real and imaginary parts explicit, at least as
 long as $\tau$ is purely imaginary.
 
 It is straightforward to match eq.~\eqref{eq:finalresult-sunrise} to the basis defined in eq.~\eqref{eq:ab_def} by matching
 the corresponding $q$-expansions. We obtain\footnote{We stress here that, since $\mathcal{M}_n(\Gamma_1(6)) \subset \mathcal{M}_n(\Gamma(6)) $, this is always possible.}
\beq\bsp
\label{eq:finalresult-sunriseAB}
 \mathcal{T}^{(0,1)}_1(t(\tau)) &\,= \frac{{\rm Cl}_2(\pi/3)}{2\pi} 
 - 2 \left[  I(\eisa_{3,6,1,0},1;\tau) - 5 I(\eisa_{3,6,1,3},1;\tau) + 5 I(\eisa_{3,6,2,3},1;\tau) \right]\,, \\
  \mathcal{T}^{(0,1)}_2(t(\tau)) &\,= 
 - \frac{i}{\pi}\left[  I(\eisa_{3,6,1,0};\tau) - 5 I(\eisa_{3,6,1,3};\tau) + 5  I(\eisa_{3,6,2,3};\tau) \right]\,,
\esp\eeq
where we have added the superscript $(0,1)$ to stress that this solution is defined for $0<t<1$. By inspecting the result, we see that
all terms in eq.~\eqref{eq:finalresult-sunriseAB} are purely real or imaginary, which guarantees that the master integrals $\mathcal{S}_1$
and $\mathcal{S}_2$ obtained through eq.~\eqref{eq:SinT} are real, as expected.
We are now ready to focus on the analytic continuation and numerical evaluation of eq.~\eqref{eq:finalresult-sunriseAB}
 for all values of $t$.

\subsection{Analytic continuation}
In the rest of this section we will study the behaviour of the solution in eq.~\eqref{eq:finalresult-sunriseAB} for different values of  $t$,
in order to produce suitable analytical representations for the two integrals, which are valid across the whole phase-space and can be evaluated numerically
by fast converging ($q$-)series expansions.

Our main goal will be to remap $\tau$ to the fundamental domain
in each region. Before doing so, we show how the homogeneous solutions in eq.~\eqref{eq:psi1_def}, and therefore $\tau$,
can be extended from the region $0<t<1$ to arbitrary (real) values of $t$. 
Indeed, the analytical expression for $\tau$ itself depends on the homogeneous solutions
of the differential equation~\eqref{eq:sunrise-deq},  which have been given explicitly in eq.~\eqref{eq:psi1_def} for $0<t<1$. 
Since eq.~\eqref{eq:sunrise-deq} has regular singular points at $t\in\{0,1,9,\pm \infty\}$, in general we expect eq.~\eqref{eq:psi1_def} to define local solutions
to the differential equations which are only valid in some region, in our case the region $0<t<1$. Hence, our first goal will be to find an analytic continuation of the local homogeneous solutions 
to all values of $t$. This can be achieved by finding suitable \emph{auxiliary} 
homogeneous solutions which are well defined in the
three remaining regions $t<0$, $1<t<9$ and $t>9$, and then appropriately matching them at the branching points with the solution determined in eq.~\eqref{eq:psi1_def}.

While this is standard material, we report here the main steps for convenience of the reader. First of all, let us define 
$$ 
\lambda_1=\frac{t_{14}(t)t_{23}(t)}{t_{13}(t)t_{24}(t)} \,, \qquad \lambda_2=\frac{t_{13}(t)t_{24}(t)}{t_{12}(t)t_{34}(t)}\,,
$$
$$
\lambda_3 = \frac{1}{2}\left( 1  - \frac{t^2-6 t - 3}{(t-1) \sqrt{t_{12}(t)t_{13}(t)}} \right)\,,
$$
where the roots are defined in eq.~\eqref{eq:SR_t_i_def}.
In order to highlight the local character of the solutions in eq.~\eqref{eq:psi1_def}, we slightly change of notation and write the local solutions as
\beq\bsp
  \Phi^{(0,1)}_1(t) & = \frac{4}{\sqrt{t_{13}t_{24}}}\,{\K}\left(\lambda_1 \right)\,,\qquad \Phi_2^{(0,1)}(t)  = i\, \frac{4}{\sqrt{t_{13}t_{24}}}\,{\K}\left(1-\lambda_1 \right)\,,
\esp\eeq
where we are using now the letter $\Phi$ instead of the letter $\Psi$ and
we added the superscript $(0,1)$ to indicate that this solution is valid only for $0<t<1$. The reason for this change of
notation will become clear soon.
In order to obtain homogeneous solutions valid for all values of $t$, we define 
three  pairs
of auxiliary functions $\Phi_i^{(a,b)}(t)$, which are all local homogeneous solutions of eq.~\eqref{eq:sunrise-deq}
and which are
well defined and purely real (or imaginary) in the corresponding region $a<t<b$.
Following ref.~\cite{Remiddi:2016gno,Bogner:2017vim} 
we can take
\beq\bsp \label{eq:functions}
 & \Phi_1^{(1,9)}(t) = \frac{4}{\sqrt{t_{14}t_{23}}}\,{\K}\left(1/\lambda_1 \right)\,,\qquad \Phi_2^{(1,9)}(t)  = i\, \frac{4 }{\sqrt{t_{14}t_{23}}}\,{\K}\left(1-1/\lambda_1 \right)\,.
\\
  & \Phi_1^{(9,\infty)}(t)  = \frac{4}{\sqrt{t_{12}t_{34}}}\,{\K}\left(\lambda_2 \right)\,,\qquad \Phi_2^{(9,\infty)}(t)  = i\, \frac{4 }{\sqrt{t_{12}t_{34}}}\,{\K}\left(1-\lambda_2 \right)\,.
\\
  & \Phi_1^{(-\infty,0)}(t) = \frac{4}{(t_{12} t_{34} t_{13}t_{24})^{1/4} }\,{\K}\left( \lambda_3\right)\,,\qquad \Phi_2^{(-\infty,0)}(t)  = i\, \frac{1}{(t_{12} t_{34} t_{13}t_{24})^{1/4}}\,{\K}\left(1-\lambda_3 \right)\,.
\esp\eeq
These functions form a complete set of local homogeneous solutions from which we can patch together the global solutions $\Psi_j(t)$
with the correct analyticity properties for all values of $t$:
\beq\bsp\label{eq:global_SR_solution}
& \begin{pmatrix}
\Psi_1(t) &
\Psi_2(t)
\end{pmatrix}
=  \begin{pmatrix}
\Phi_1^{(a,b)}(t) &
\Phi_2^{(a,b)}(t)
\end{pmatrix} M^{(a,b)} \, \quad \mbox{for} \quad a<t<b\,,
\esp\eeq
where $M^{(a,b)}$ are $2 \times 2$ matrices with numerical entries.
By definition, we clearly must have 
$$M^{(0,1)} = \begin{pmatrix} 1 & 0 \\ 0 & 1 \end{pmatrix}\,, $$
which allows us to recover the original definition in eq.~\eqref{eq:psi1_def} for $0<t<1$.
The remaining matrices can be found by imposing that
the homogeneous solution is analytic in the neighbourhood of each regular singular point $t=a$ (with possibly logarithmic singularities at $t=a$).
With our choice of functions $\Phi_i^{(a,b)}$ they read
\begin{equation}
M^{(1,9)} = \begin{pmatrix} 1 & 0 \\ 3 & 1 \end{pmatrix}\,, \quad 
M^{(9,\infty)} = \begin{pmatrix} -2 & -1 \\ 3 & 1 \end{pmatrix}\,, \quad
M^{(-\infty,0)} = \begin{pmatrix} 1 & 1/2 \\ 0 & 2 \end{pmatrix}\,. 
\end{equation}
Once the continued homogenous solutions 
$\Psi_j$ have been determined, the complete matrix of solutions $\mathcal{W}_S(t)$ of eq.~\eqref{eq:sunrise_wronskian} 
can be extended from the region $0<t<1$ to all real values of $t$ by simply using the piecewise 
definition of $\Psi_j(t)$ of eq.~\eqref{eq:global_SR_solution} in eq.~\eqref{eq:sunrise_wronskian}. In the same way,
we can extend the definition of $\tau(t)$ in eq.~\eqref{eq:psi3def} to all values of $t$.

\begin{figure}[!th]
\begin{center}
\includegraphics[scale=0.6]{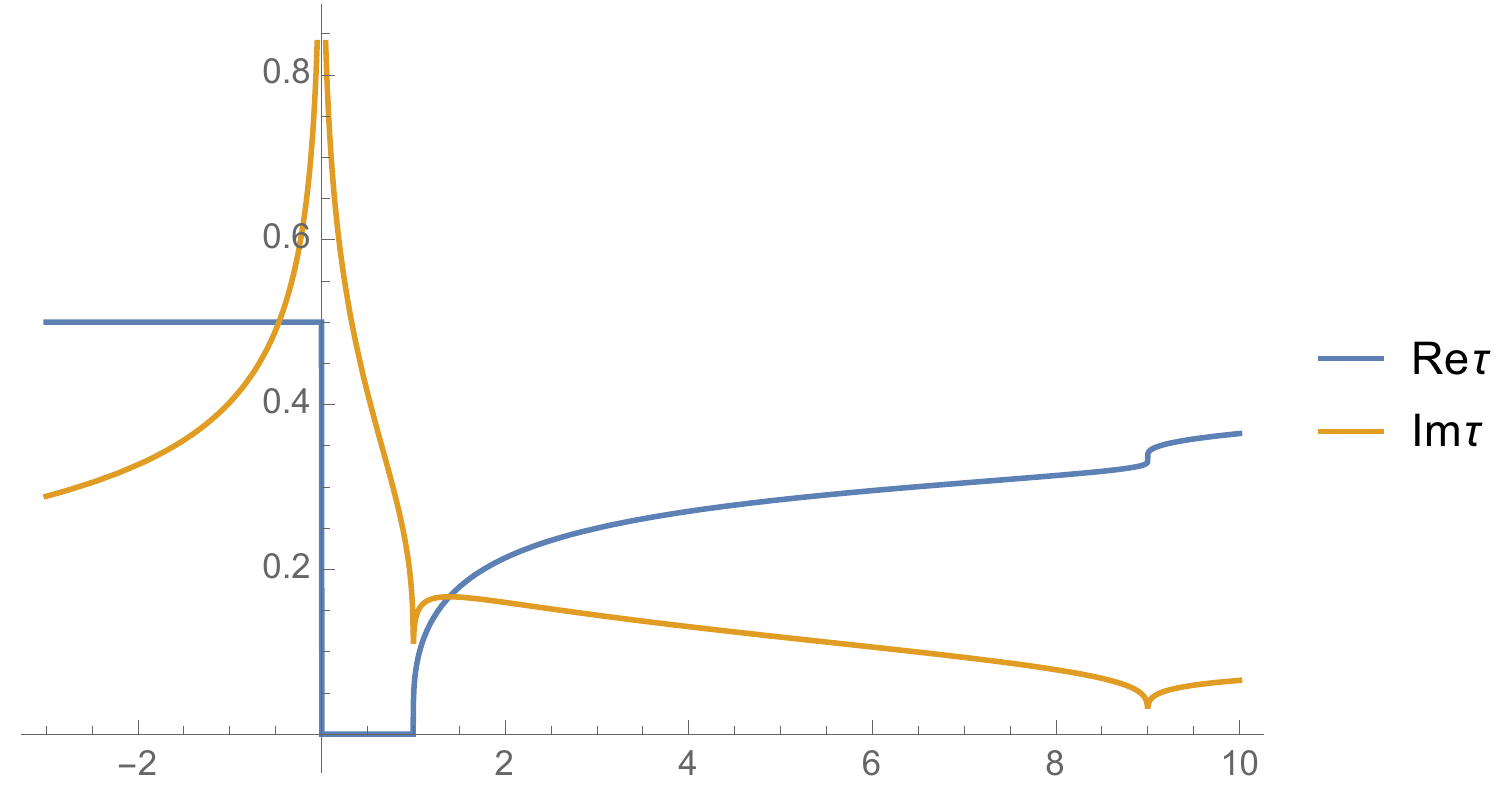}
\caption{\label{fig:tausun}The real and imaginary parts of $\tau$.}
\end{center}
\end{figure}

\begin{figure}[!th]
\begin{center}
\includegraphics[scale=0.6]{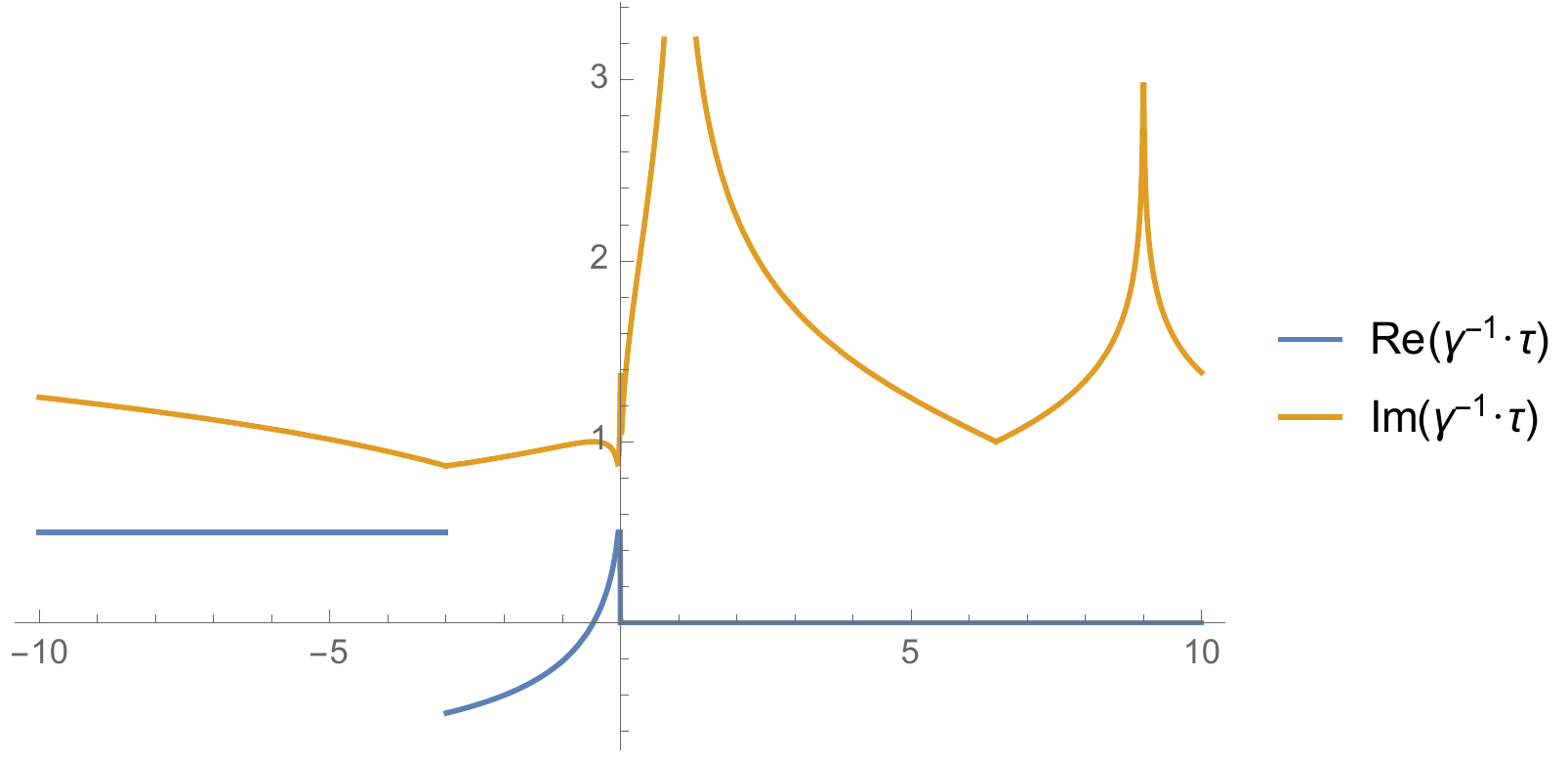}
\caption{\label{fig:tau_remapped}The real and imaginary parts of $\gamma_j^{-1} \cdot \tau^{(a,b)}$.}
\end{center}
\end{figure}

With our correctly continued homogeneous solutions, we can study when we need to
perform a modular transformation 
to remap $\tau$ to the fundamental domain and, in this way, accelerate the convergence of the
$q$-expansion of the corresponding iterated Eisenstein integrals.
It is useful to start by plotting  $\tau$ as function of $t$, see fig.~\ref{fig:tausun}.
It is easy to see that our definition for $\tau$ almost never lies in the fundamental domain, except for a small
region around the point $t=0$.
As we have seen in the case of the ${}_2F_1$ hypergeometric function, one can find a fast-converging representation in the region $a<t<b$ via a 
modular transformation such that $\tau$ is mapped to the fundamental domain.
%
In order to 
achieve this in general we must use different modular transformations, depending on the value of $t$. We find in particular that we 
need to
consider 8 different transformations $\gamma_t \in \SL(2,\mathbb{Z})$, 
such that $\gamma_t^{-1} \cdot \tau(t)$ lies
in the fundamental domain. Defining the four special points
\beq\bsp
&r_1 = -3\,,  \quad r_2 = 5 - 2^{1/3}\, 4 \,, \quad r_3 = 3 + 2 \sqrt{3}  ( 1 - 3^{1/4} \sqrt{2} )\,, \\
& r_4 = 3 + 2 \sqrt{3}\,, \quad r_5 = 3 + 2 \sqrt{3}  ( 1 + 3^{1/4} \sqrt{2} )\,,
\esp\eeq
we find
\beq\label{eq:gamma_t_def}
\gamma_t=\left\{\begin{array}{ll}
\gamma_{0a} \,,& \textrm{ if } t < r_1 \,,\\
\gamma_{0b} \,,& \textrm{ if } r_1 \le t < r_2\,,\\
\gamma_{1a} \,,& \textrm{ if } r_2<t< r_3\,,\\
\gamma_{1b} \,,& \textrm{ if } r_3 < t < 1\,,\\
\gamma_{2a} \,,& \textrm{ if } 1 \le t < r_4\,,\\
\gamma_{2b} \,,& \textrm{ if }  r_4 \le t < 9\,,\\
\gamma_{3a} \,,& \textrm{ if }  9 \le t < r_5\,,\\
\gamma_{3b} \,,& \textrm{ if } t \ge r_5 \,,
\end{array}\right.
\eeq
with
\beq\bsp
&\gamma_{0a} \,=\mat{-1&1}{-2&1}\,, \quad \gamma_{0b} \,=\mat{0&-1}{1&-1}\,, \quad
\gamma_{1a} \,=\mat{1&0}{0&1}\,, \quad \gamma_{1b} \,=\mat{0&-1}{1&0}\,, \\
&\gamma_{2a} \,=\mat{0&-1}{1&-3}\,, \quad \gamma_{2b} \,=\mat{-1&0}{-3&-1}\,, \quad
\gamma_{3a} \,=\mat{-1&1}{-3&2}\,, \quad \gamma_{3b} \,=\mat{-1&-1}{-2&-3}\,.
\esp\eeq
Recalling the expression for the $j$-invariant in eq.~\eqref{eq:jinv_sun}, we see that the apparently
random special points  $t=r_i$, $i-1,..,5$
where the $\gamma_t \in \SL(2,\mathbb{Z})$ is discontinuous, 
are always points $t_i$ such that $j(t_i) \in\{0,1,\infty\}$, similar to what was observed for
the ${}_2 F_1$ function. It is important to stress that these regions do not always overlap
with the regions delimited by the regular singular points $t_i \in \{0,1,9,\infty\}$, in particular
the region $1a$ lies across the point $t=0$. This is not a problem. It just tells us that as long as $t$ varies in the small region 
$r_2<t< r_3$, i.e.
$-0.0397...\leq t < 0.0167... $, the relevant solutions (either the ones valid in the range $(0,1)$ or the ones valid in the range $(-\infty,0)$) can be taken
as they are, without the need of remapping the corresponding $\tau$ to the fundamental domain.
By defining $\tau_j = \gamma_j^{-1} \cdot \tau$, we can plot the remapped
value in function of $t$ and check that it indeed always lies in the fundamental domain, see fig.~\ref{fig:tau_remapped}.

With the matrices defined in eq.~\eqref{eq:gamma_t_def}, we can now perform the relevant
transformations on the iterated integrals in eq.~\eqref{eq:finalresult-sunriseAB} and obtain
new combinations of integrals evaluated in the fundamental domain. Clearly, there is nothing to do
in region $1a$, as $\gamma_{1a} = \mathds{1} $. The first non-trivial transformation is in region $1b$, i.e.
for $r_3<t<1$
where the transformation induced by $\gamma_{1b} \in \SL(2,\mathbb{Z})$  has to applied in order to map $\tau$ back to the fundamental
domain.
We define $\tau = \gamma_{1b} \cdot \tau_{1b}$, and apply the algorithms described in this paper
to obtain
\begin{align}
\label{eq:finalresult-sunriseAB_1}
 \mathcal{T}^{(1b)}_1(t(\tau_{1b})) &\,= \frac{1}{\pi} \left[  I(\eisb_{3,6,0,1};\tau_{1b}) - 5 I(\eisb_{3,6,3,1};\tau_{1b}) 
 + 5 I(\eisb_{3,6,3,2};\tau_{1b}) \right] \\
  \mathcal{T}^{(1b)}_2(t(\tau_{1b})) &\,= 
 - \frac{i}{8}\left[ \pi - 16 I(\eisb_{3,6,0,1},1;\tau_{1b}) + 80 I(\eisb_{3,6,3,1},1;\tau_{1b}) - 80  I(\eisb_{3,6,3,2},1;\tau_{1b}) \right]
 \,,\nonumber
\end{align}
where now $\tau_{1b}$ lies in the fundamental domain. The analytic expressions for the master integrals
in the remaining regions can be found in Appendix~\ref{app:sun_results}.

\begin{figure}[!th]
\centering
\begin{subfigure}{.5\textwidth}
  \centering
\includegraphics[scale=0.48]{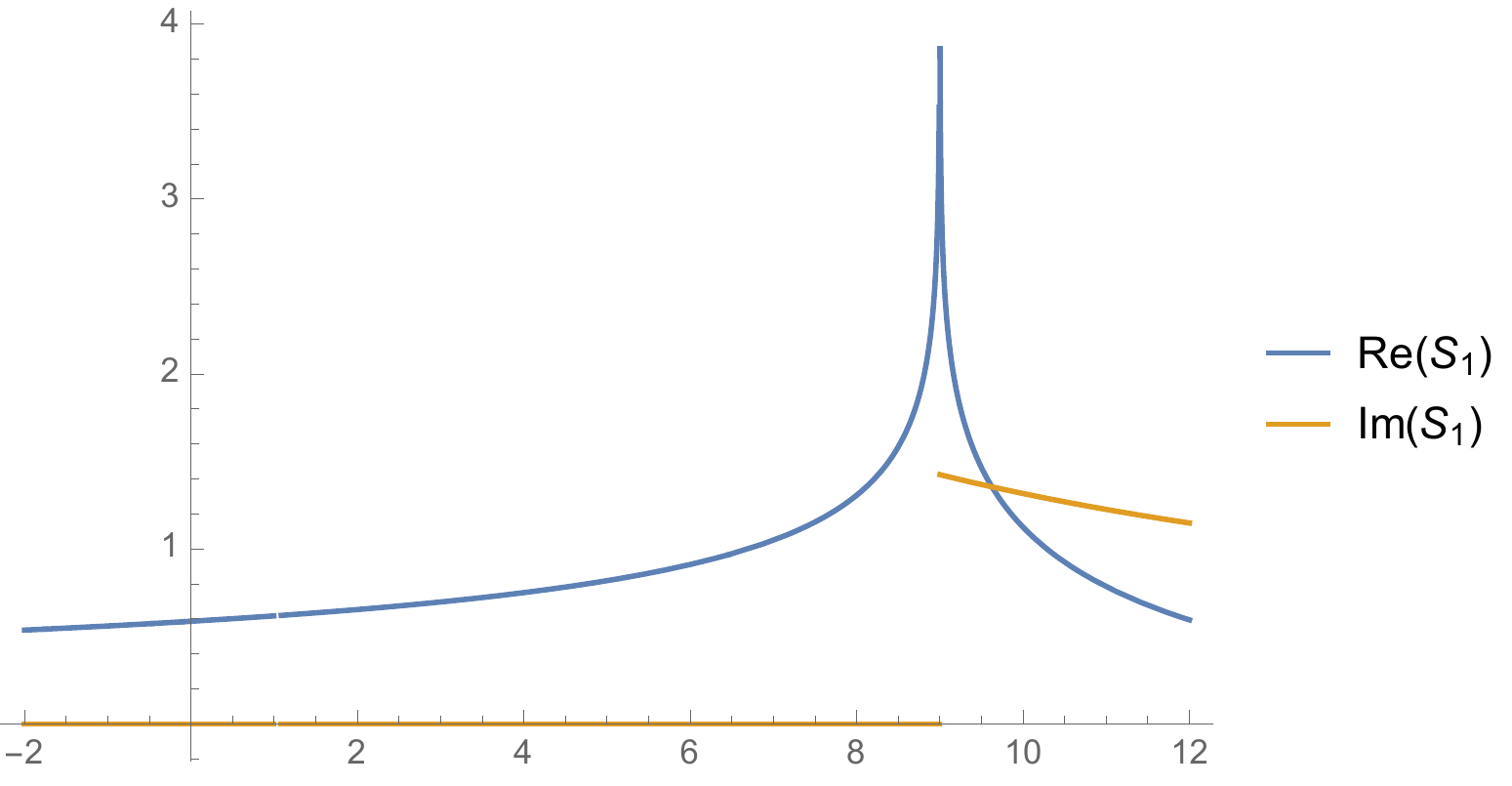}
\caption{The master integral $\mathcal{S}_1$.}
  \label{fig:sub1}
\end{subfigure}%
\begin{subfigure}{.5\textwidth}
  \centering
\includegraphics[scale=0.48]{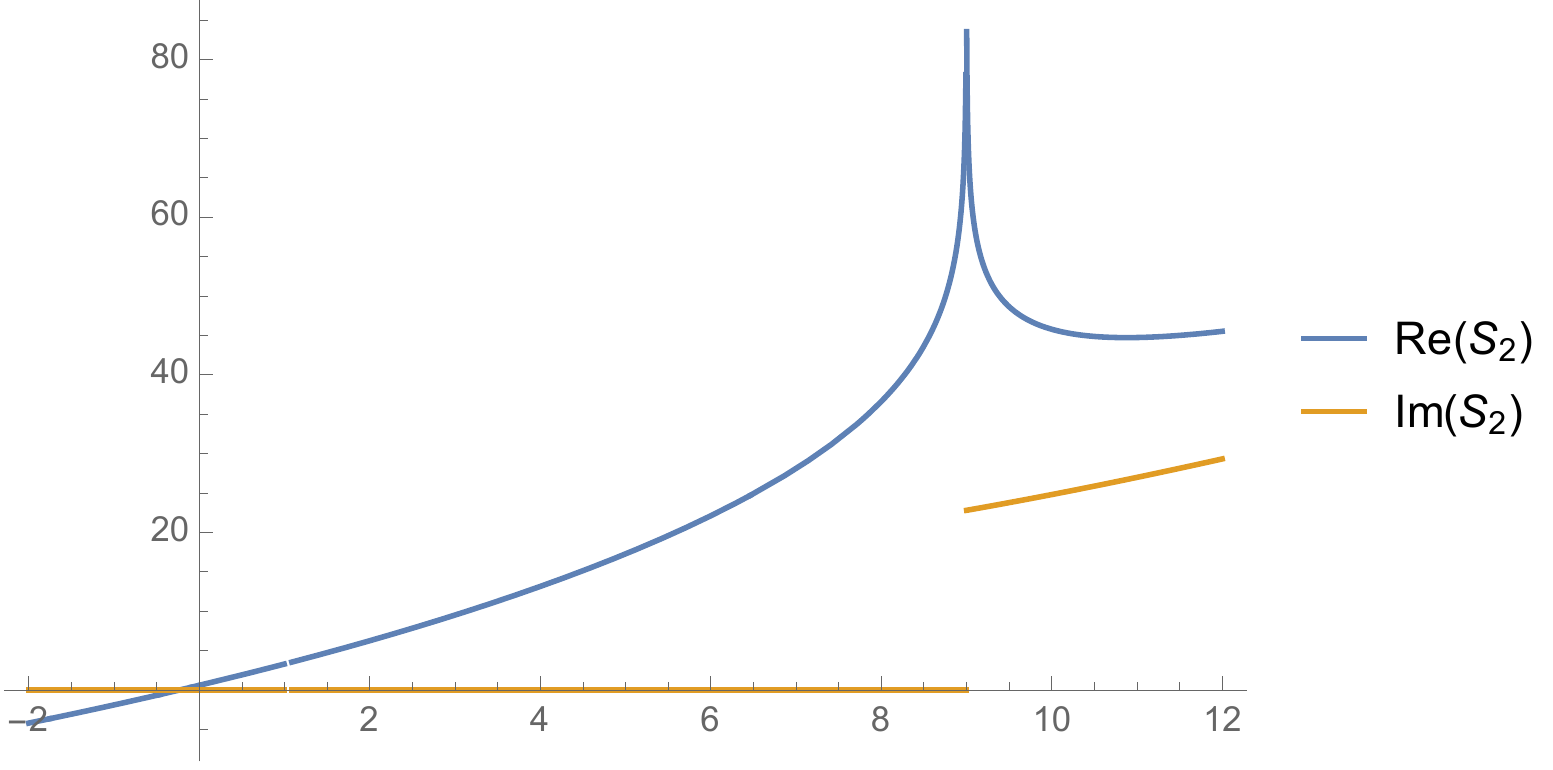}
\caption{The master integral $\mathcal{S}_2$.}
  \label{fig:sub2}
\end{subfigure}
\caption{The real and imaginary parts of the two master integrals of the sunrise graph.}
\label{fig:plotsun}
\end{figure}

We can then use these expressions to plot the result for the original master integrals $\mathcal{S}_1$
and $\mathcal{S}_2$ as a function of $t$. Since the result is now always evaluated
for $\tau(t)$ in the fundamental domain, we can get very precise numerical evaluations by 
considering only the first few terms of the corresponding $q$-series expansions.
For example, by keeping $10$ terms of the expansion and running for a few seconds on 
a laptop computer we can get the plots in fig.~\ref{fig:plotsun}.
Our numerical results have been checked thoroughly against the known analytical results in ref.~\cite{Remiddi:2016gno}
and the numerical program SecDec~\cite{Borowka:2015mxa,Borowka:2017idc}.
As a concluding remark, we want to point out that all techniques used here can be straightforwardly applied to 
analytically continue and numerically evaluate other Feynman diagrams that can be expressed in terms of the same
class of functions. Obvious examples are given by the so-called Kite integral~\cite{Remiddi:2016gno,Bogner:2017vim} and the three-loop 
equal-mass banana graph~\cite{Bloch:2014qca,Primo:2017ipr,Broedel:2019kmn}.

%% file: conclusion.tex

\section{Conclusion}
\label{sec:conclusion}

In the last years it has started to become clear that  multiloop Feynman integrals in dimensional regularisation 
can be naturally expressed in terms of iterated integrals of sets of differential forms defined on increasingly complicated complex (hyper)-surfaces.
The simplest case is represented by multiple polylogarithms, which are defined as iterated integrals of rational functions on the Riemann sphere.
The next-to-simplest case is instead constituted by their generalisation to genus-one Riemann surfaces -- tori, or equivalently to elliptic curves --
and have been dubbed elliptic multiple polylogarithms. A subset of the latter, which has found wide application in a multitude of physically
interesting problems which depend on one single ratio of kinematical invariants, are closely related to iterated integrals of Eisenstein series.

While a lot is known about the geometrical aspects of the construction of eMPLs and iterated Eisenstein integrals, 
their use in physically realistic problems
requires being able to control the details of their branch cut structure (i.e. their analytic continuation) and to devise strategies
to evaluate them numerically with high-precision for any values of their arguments.
This problem can be solved in full generality for the simple case of MPLs, but till today still a little was known about how to 
treat these new classes of functions.
In this paper a first step in this direction has been made. We have presented a 
set of algorithms which allow one to analytically continue iterated integrals of Eisenstein series
to any region of the phase-space, in a way that makes it straightforward to evaluate them numerically
with very high-precision through fast converging $q$-series expansions. We have shown the techniques in detail by applying them
to a mathematical problem, i.e. the Taylor expansion of a class of hypergeometric functions, and to the physical problem of the 
calculation of the two-loop massive sunrise graph.
We expect generalisations of the algorithms described here to be able to address the evaluation of more 
general eMPLs.

%% file: app_SL2Z.tex

\section{The modular group $\SL(2,\mathbb{Z})$}
\label{app:SL2Z}
In this appendix we review some useful algorithms related to the modular group $\SL(2,\mathbb{Z})$. We start by an algorithm to decompose any element of $\SL(2,\mathbb{Z})$ into a finite product of the generators $S$, $T$ and $-\mathds{1}$.
In a second subsection we discuss, for some $\tau\in\mathbb{H}$, how to find $\gamma\in\SL(2,\mathbb{Z})$ such that $\gamma\cdot \tau$ lies in the fundamental domain for $\SL(2,\mathbb{Z})$.

\subsection{Decomposing elements into a product of $S$'s  and $T$'s}
Consider $\gamma=\mat{\gamma_{11}&\gamma_{12}}{\gamma_{21}&\gamma_{22}}\in\SL(2,\mathbb{Z})$. We start by describing what happens if we multiply $\gamma$ from the left by $S$ or $T$. If $N$ is any integer, we have
\beq
S\gamma = \mat{-\gamma_{21}&-\gamma_{22}}{\gamma_{11}&\gamma_{12}}\textrm{~~and~~} T^N\gamma = \mat{\gamma_{11}+N\gamma_{21}&\gamma_{12}+N\gamma_{22}}{\gamma_{21}&\gamma_{22}}\,.
\eeq
We now describe the algorithm to decompose $\gamma$ into a product of generators.
\begin{enumerate}
\item[-]{\bf Step 1:} If $|\gamma_{11}|<|\gamma_{21}|$, replace $\gamma$ by
\beq
\gamma = \mat{\gamma_{11}&\gamma_{12}}{\gamma_{21}&\gamma_{22}}\longrightarrow S\gamma =  \mat{-\gamma_{21}&-\gamma_{22}}{\gamma_{11}&\gamma_{12}}\,.
\eeq
After this step, we can assume that $|\gamma_{11}|>|\gamma_{21}|$.
\item[-]{\bf Step 2:} If $\gamma_{21} = 0$, go to Step 3. Otherwise, we use Euclidean division to write
\beq
\gamma_{11} = \gamma_{21}\,q + r\,,\qquad 0\le r<|\gamma_{21}|\,,
\eeq
and we replace $\gamma$ by
\beq
\gamma\longrightarrow ST^{-q}\gamma =  \mat{-\gamma_{21}&-\gamma_{22}}{\gamma_{11}-q\gamma_{21}&\gamma_{12}-q\gamma_{22}}=  \mat{-\gamma_{21}&-\gamma_{22}}{r&\gamma_{12}-q\gamma_{22}}\,.
\eeq
If $r\neq 0$, return to Step 2. At each iteration the value of $r$ is strictly decreasing, and since $r>0$, it will eventually reach 0.
\item[-]{\bf Step 3:} The matrix now has the form $\gamma=\mat{\gamma_{11}&\gamma_{12}}{0&\gamma_{22}}$. Since $\det\gamma=1$, we must have $\gamma_{11}=\gamma_{22}=\pm 1$. Replace $\gamma$ by
\beq
\gamma\longrightarrow T^{-\gamma_{11}\gamma_{12}}\gamma = \mat{\gamma_{11} &0}{0&\gamma_{22}} = \mat{\pm1&0}{0&\pm1}.
\eeq
\end{enumerate}
At the end of this procedure we have constructed a product of $S$'s and $T$'s which is the inverse of our original $\gamma$ (up to a sign). This easily fournishes the desired decomposition into the generators $S$ and $T$, using the fact that $S^{-1}=S$ and $(T^N)^{-1} = T^{-N}$.

\subsection{Mapping a point in the upper half-plane to the fundamental domain}
In this appendix we describe how to find for every $\tau\in\mathbb{H}$ an element $\gamma\in\SL(2,\mathbb{Z})$ such that $\gamma\cdot \tau$ lies in the fundamental domain $\cF$ defined in eq.~\eqref{eq:fundamental_domain}. We will construct $\gamma$ iteratively using the following algorithm:

\begin{enumerate}
\item[-]{\bf Step 1:} We can easily find an integer $N$ such that $-\frac{1}{2}\le\textrm{Re } \tau+N<\frac{1}{2}$. Replace $\tau$ by
\beq
\tau\longrightarrow T^N\cdot \tau=\tau+N\,.
\eeq
\item[-] {\bf Step 2:} If $|\tau|>1$, we are done. Otherwise, replace $\tau$ by
\beq
\tau\longrightarrow S\cdot \tau = -1/\tau\,.
\eeq
If after this replacement $\tau\in\cF$, we are done. Otherwise we return to Step 1. At each iteration the imaginary part of $\tau$ is strictly increasing, and therefore the algorithm will eventually terminate.
\end{enumerate}

%% file: app_2f1_results.tex

\section{Results for the elliptic ${}_2F_1$ function}
\label{app:2f1_results}
In this appendix we present the results for the two master integrals $T_1$ and $T_2$ in eq.~\eqref{eq:2F1_masters}, in each of the six regions defined in eq.~\eqref{eq:gamma_z_def}. In each region the master integrals can be written as (cf.~eq.~\eqref{eq:T_S_U})
\beq
\left(\begin{array}{c}
T_1(z)\\ T_2(z)\end{array}\right) = S_{j\alpha}(z)
\left(\begin{array}{c}
U_1^{{j\alpha}}(\tau_{j\alpha})\\ U_2^{{j\alpha}}(\tau_{j\alpha})\end{array}\right)\,,\qquad j\in\{1,2,3\}\,,\,\,\alpha\in\{a,b\}\,.
\eeq
In the following we list the matrices $S_{j\alpha}(z)$ and the first few terms in the $\eps$-expansion of the functions $U_i^{{j\alpha}}(\tau_{j\alpha})$ for each region.

\beq
S_{1a}(z) = \left(
\begin{array}{cc}
 \frac{2 i \K\left(\frac{1}{1-z}\right)}{\sqrt{1-z}} & 0 \\
 -\frac{2 i \sqrt{1-z} \E\left(\frac{1}{1-z}\right)}{z (1+6 \epsilon )}-\frac{2 i (2 z-1) \K\left(\frac{1}{1-z}\right)}{3 \sqrt{1-z} z (1+6 \epsilon)} & -\frac{\pi  \sqrt{1-z}}{z (1+6 \epsilon) \K\left(\frac{1}{1-z}\right)} \\
\end{array}
\right)\,.
\eeq

\beq\bsp
U_1^{1a}(\tau_{1a}) &\,=
-2i\pi I(1;\tau_{1a})+i\pi\,\epsilon\left[1-2\log\left(1-\frac{1}{z}\right)I(1;\tau_{1a})\right]\\
&\,+i\pi\,\epsilon^2\Bigg[-360I(1,\eisa_{4,2,0,0},1;\tau_{1a})+\left(\frac{4\pi^2}{3}-\log^2\left(1-\frac{1}{z}\right)\right)I(1;\tau_{1a})\\
&\,+\log\left(1-\frac{1}{z}\right)-\frac{30}{\pi^2}\,\zeta_3\Bigg]+\ord(\eps^3)
\,.
\esp\eeq

\beq\bsp
U_2^{1a}(\tau_{1a}) &\,=
-1-\epsilon\,\log\left(1-\frac{1}{z}\right)+\epsilon^2\,\Bigg[-180I(\eisa_{4,2,0,0},1;\tau_{1a})-\frac{1}{2}\log^2\left(1-\frac{1}{z}\right)\\
&\,+\frac{2\pi^2}{3}\Bigg]
+\ord(\eps^3)\,.
\esp\eeq

\beq
S_{1b}(z) = 
\left(
\begin{array}{cc}
 \frac{2 \K\left(\frac{z}{z-1}\right)}{\sqrt{1-z}} & 0 \\
 \frac{2 \sqrt{1-z} \E\left(\frac{z}{z-1}\right)}{z (1+6 \epsilon)}-\frac{2 (2-z) \K\left(\frac{z}{z-1}\right)}{3 \sqrt{1-z} z (1+6 \epsilon )} & -\frac{i \pi  \sqrt{1-z}}{z (1+6 \epsilon) \K\left(\frac{z}{z-1}\right)} \\
\end{array}
\right)
\,.
\eeq

\beq\bsp
U_1^{1b}(\tau_{1b}) &\,=1+\epsilon\,\Bigg[\log\left(1-\frac{1}{z}\right)-2\pi^2I(1;\tau_{1b})\Bigg]+\epsilon^2\,\Bigg[180I(1,\eisa_{4,2,0,0};\tau_{1b})\\
&\,-2\pi^2\log\left(1-\frac{1}{z}\right)I(1;\tau_{1b})+\frac{1}{2}\log^2\left(1-\frac{1}{z}\right)+\frac{\pi^2}{6}\Bigg]
+\ord(\eps^3)\,.
\esp\eeq

\beq\bsp
U_2^{1b}(\tau_{1b}) &\,=
i\pi\,\epsilon+i\pi\,\epsilon^2\Bigg[\log\left(1-\frac{1}{z}\right)-\frac{90}{\pi^2}\,I(\eisa_{4,2,0,0};\tau_{1b})\Bigg]
+\ord(\eps^3)\,.
\esp\eeq

\beq
S_{2a}(z) = 
\left(
\begin{array}{cc}
 2 \K(z) & 0 \\
 \frac{2 \E(z)}{z (1+6 \epsilon)}-\frac{2 (2-z) \K(z)}{3 z (1+6 \epsilon)} & -\frac{i \pi }{z (1+6 \epsilon) \K(z)} \\
\end{array}
\right)
\,.
\eeq

\beq\bsp
U_1^{2a}(\tau_{2a}) &\,=
1+\epsilon\,\Bigg[\log\left(\frac{1-z}{z}\right)-2\pi^2I(1;\tau_{2a})\Bigg]+\epsilon^2\,\Bigg[180I(1,\eisa_{4,2,0,0};\tau_{2a})\\
&\,-2\pi^2\log\left(\frac{1-z}{z}\right)I(1;\tau_{2a})+\frac{1}{2}\log^2\left(\frac{1-z}{z}\right)+\frac{\pi^2}{6}\Bigg]
+\ord(\eps^3)\,.
\esp\eeq

\beq\bsp
U_2^{2a}(\tau_{2a}) &\,=
i\pi\,\epsilon+i\pi\,\epsilon^2\,\Bigg[\log\left(\frac{1-z}{z}\right)-\frac{90}{\pi^2}\,I(\eisa_{4,2,0,0};\tau_{2a})\Bigg]
+\ord(\eps^3)\,.
\esp\eeq

\beq
S_{2b}(z) = 
\left(
\begin{array}{cc}
 -2 i \K(1-z) & 0 \\
 \frac{2 i \E(1-z)}{z (1+6 \epsilon )}-\frac{2 i (z+1) \K(1-z)}{3 z \
(1+6 \epsilon )} & \frac{\pi }{z (1+6 \epsilon ) \K(1-z)} \\
\end{array}
\right)
\,.
\eeq

\beq\bsp
U_1^{2b}(\tau_{2b}) &\,=
+2i\pi I(1;\tau_{2b})+i\pi\,\epsilon\,\Bigg[2\log\left(\frac{1-z}{z}\right)I(1;\tau_{2b})-1\Bigg]\\
&\,+i\pi\,\epsilon^2\,\Bigg[360I(1,\eisa_{4,2,0,0},1;\tau_{2b})+\left(\log^2\left(\frac{1-z}{z}\right)-\frac{4i\pi^3}{3}\
\right)I(1;\tau_{2b})\\
&\,-\log\left(\frac{1-z}{z}\right)+\frac{30}{\pi^2}\,\zeta_3\Bigg]
+\ord(\eps^3)\,.
\esp\eeq

\beq\bsp
U_2^{2b}(\tau_{2b}) &\,=
1+\epsilon\,\log\left(\frac{1-z}{z}\right)+\epsilon^2\,\Bigg[180I(\eisa_{4,2,0,0},1;\tau_{2b})+\frac{1}{2}\log^2\left(\frac{1-z}{z}\right)\\
&\,-\frac{2\pi^2}{3}\Bigg]
+\ord(\eps^3)\,.
\esp\eeq

\beq
S_{3a}(z) = 
\left(
\begin{array}{cc}
 \frac{2 i \K\left(\frac{z-1}{z}\right)}{\sqrt{z}} & 0 \\
 -\frac{2 i \E\left(\frac{z-1}{z}\right)}{\sqrt{z} (1+6 \epsilon )}+\frac{2 i (z+1) \K\left(\frac{z-1}{z}\right)}{3 z^{3/2} (1+6 \epsilon )} & -\frac{\pi }{\sqrt{z} (1+6 \epsilon ) \K\left(\frac{z-1}{z}\right)} \\
\end{array}
\right)
\,.
\eeq

\begin{align}
\nonumber U_1^{3a}(\tau_{3a}) &\,=
1-2i\pi I(1;\tau_{3a})+\epsilon\,\Bigg[\log\left(1-\frac{1}{z}\right)-\left(2\pi^2+2i\pi\log\left(1-\frac{1}{z}\right)\right)I(1;\tau_{3a})\Bigg]\\
&\,+\epsilon^2\,\Bigg[180I(1,\eisa_{4,2,0,0};\tau_{3a})-360i\pi I(1,\eisa_{4,2,0,0},1;\tau_{3a})\\
\nonumber&\,-\left(i\pi\log^2\left(1-\frac{1}{z}\right)+2\pi^2\log\left(1-\frac{1}{z}\right)+\frac{10i\pi^3}{3}\right)I(1;\tau_{3a})+\frac{1}{2}\log^2\left(1-\frac{1}{z}\right)\\
\nonumber&\,-\frac{\pi^2}{3}-\frac{90i}{3\pi}\,\zeta_3\Bigg]
+\ord(\eps^3)\,.
\end{align}

\begin{align}
\nonumber U_2^{3a}(\tau_{3a}) &\,=
-1+\epsilon\Bigg[i\pi -\log\left(1-\frac{1}{z}\right)\Bigg]+\epsilon^2\,\Bigg[-\frac{90i}{\pi}\,I(\eisa_{4,2,0,0};\tau_{3a})-180I(\eisa_{4,2,0,0},1;\tau_{3a})\\
&\,-\frac{1}{2}\log^2\left(1-\frac{1}{z}\right)+i\pi\log\left(1-\frac{1}{z}\right)+\frac{5\pi^2}{3}\Bigg]
+\ord(\eps^3)\,.
\end{align}

\beq
S_{3b}(z) = 
\left(
\begin{array}{cc}
 -\frac{2 \K\left(\frac{1}{z}\right)}{\sqrt{z}} & 0 \\
 \frac{2 (2 z-1) \K\left(\frac{1}{z}\right)}{3 z^{3/2} (1+6 \epsilon )}-\frac{2 \E\left(\frac{1}{z}\right)}{\sqrt{z} (1+6 \epsilon )} & \frac{i \pi }{\sqrt{z} (1+6 \epsilon ) \K\left(\frac{1}{z}\right)} \\
\end{array}
\right)
\,.
\eeq

\begin{align}
\nonumber U_1^{3b}(\tau_{3b}) &\,= -1-2i\pi I(1;\tau_{3b}) +\epsilon\,\Bigg[-2i\pi\log\left(1-\frac{1}{z}\right)I(1;\tau_{3b})-\log\left(1-\frac{1}{z}\right)+i\pi\Bigg]\\
&\,+
\epsilon^2\,\Bigg[-180I(1,\eisa_{4,2,0,0};\tau_{3b})-360i\pi I(1,\eisa_{4,2,0,0},1;\tau_{3b})\\
\nonumber&\,+\left(\frac{7i\pi^3}{3}-i\pi\log^2\left(1-\frac{1}{z}\right)\right)I(1;\tau_{3b})-\frac{1}{2}\log^2\left(1-\frac{1}{z}\right)+i\pi\log\left(1-\frac{1}{z}\right)\\
\nonumber&\,+\frac{5\pi^2}{6}-\frac{180i}{6\pi}\,\zeta_3\Bigg]
+\ord(\eps^3)\,.
\end{align}

\beq\bsp
U_2^{3b}(\tau_{3b}) &\,=-1-\epsilon\,\log\left(1-\frac{1}{z}\right)
\epsilon^2\,\Bigg[\frac{90i}{\pi}\,I(\eisa_{4,2,0,0};\tau_{3b})-180I(\eisa_{4,2,0,0},1;\tau_{3b})\\
&\,-\frac{1}{2}\log^2\left(1-\frac{1}{z}\right)+\frac{7\pi^2}{6}\Bigg]
+\ord(\eps^3)\,.
\esp\eeq

%% file: app_sun_results.tex

\section{Results for the sunrise graph}
\label{app:sun_results}

We write here the analytical expressions for the continued sunrise integrals in the 8 regions defined in eq.~\eqref{eq:gamma_t_def}.
For each region $r=\{0a,0b,1a,1b,2a,2b,3a,3b\}$, we define $\tau = \gamma_r \cdot \tau_r$ and we find:

\beq\bsp
\label{eq:finalresult-sunriseAB_0a}
 \mathcal{T}^{(0a)}_1(t(\tau_{0a})) &\,=   
 \frac{i (\tau_{0a} -1)}{\pi } \Big[
5 I(\eisa_{3,6,1,1}; \tau_{0a} )-I(\eisa_{3,6,1,2}; \tau_{0a} )-5
   I(\eisa_{3,6,2,1}; \tau_{0a} ) \\ &+5 i I(\eisb_{3,6,1,1}; \tau_{0a} )+i I(\eisb_{3,6,1,2}; \tau_{0a} )+5 i
   I(\eisb_{3,6,2,1}; \tau_{0a} ) 
   \Big] \\
   &+10 I(1,\eisa_{3,6,1,1}; \tau_{0a} )-2
   I(1,\eisa_{3,6,1,2}; \tau_{0a} )-10 I(1,\eisa_{3,6,2,1}; \tau_{0a} ) \\&+2 i \Big[ 5
   I(1,\eisb_{3,6,1,1}; \tau_{0a} )+I(1,\eisb_{3,6,1,2}; \tau_{0a} )+5 I(1,\eisb_{3,6,2,1}; \tau_{0a}
   )\Big] + \frac{i \pi }{8}\,,
  \\
  \mathcal{T}^{(0a)}_2(t(\tau_{0a})) &\,= 
  \frac{(2 \tau_{0a} -1)}{\pi } \Big[-5 i I(\eisa_{3,6,1,1}; \tau_{0a} )+i I(\eisa_{3,6,1,2}; \tau_{0a} )+5 i
   I(\eisa_{3,6,2,1}; \tau_{0a} ) \\
   &+5 I(\eisb_{3,6,1,1}; \tau_{0a} )+I(\eisb_{3,6,1,2}; \tau_{0a} )+5
   I(\eisb_{3,6,2,1}; \tau_{0a} )\Big] \\
   &-20 I(1,\eisa_{3,6,1,1}; \tau_{0a} )+4
   I(1,\eisa_{3,6,1,2}; \tau_{0a} )+20 I(1,\eisa_{3,6,2,1}; \tau_{0a} )  \\
   &-4 i \Big[5
   I(1,\eisb_{3,6,1,1}; \tau_{0a} )+I(1,\eisb_{3,6,1,2}; \tau_{0a} )+5 I(1,\eisb_{3,6,2,1}; \tau_{0a}
   )\Big] -\frac{i \pi }{12} \,,
\esp\eeq

\beq\bsp
\label{eq:finalresult-sunriseAB_0b}
 \mathcal{T}^{(0b)}_1(t(\tau_{0b})) &\,=  \frac{1}{\pi } \Big[i I(\eisa_{3,6,1,1}; \tau_{0b})-5 i I(\eisa_{3,6,1,2}; \tau_{0b})
 +5 i I(\eisa_{3,6,2,1}; \tau_{0b}) \\
   &+I(\eisb_{3,6,1,1}; \tau_{0b})+5
   I(\eisb_{3,6,1,2}; \tau_{0b})-5 I(\eisb_{3,6,2,1}; \tau_{0b}) \big]\,,\\
  \mathcal{T}^{(0b)}_2(t(\tau_{0b})) &\,=  
   \frac{(\tau_{0b}-1)}{\pi } \Big[i I(\eisa_{3,6,1,1}; \tau_{0b})-5 i I(\eisa_{3,6,1,2}; \tau_{0b})+5 i
   I(\eisa_{3,6,2,1}; \tau_{0b}) \\
   &+I(\eisb_{3,6,1,1}; \tau_{0b})+5
   I(\eisb_{3,6,1,2}; \tau_{0b})-5 I(\eisb_{3,6,2,1}; \tau_{0b})
   \Big] \\
   &+2
   I(1,\eisa_{3,6,1,1}; \tau_{0b})-10 I(1,\eisa_{3,6,1,2}; \tau_{0b})+10
   I(1,\eisa_{3,6,2,1}; \tau_{0b})  \\
   &-2 i \Big[ I(1,\eisb_{3,6,1,1}; \tau_{0b})+5
   I(1,\eisb_{3,6,1,2}; \tau_{0b})-5 I(1,\eisb_{3,6,2,1}; \tau_{0b})\Big] -\frac{i \pi }{8} \,,
  \esp\eeq
  
  \beq\bsp
\label{eq:finalresult-sunriseAB_2a}
 \mathcal{T}^{(2a)}_1(t(\tau_{2a})) &\,=  \frac{1}{\pi }\Big[-5 I(\eisb_{3,6,0,1}; \tau_{2a})+I(\eisb_{3,6,3,1}; \tau_{2a})+5
   I(\eisb_{3,6,3,2}; \tau_{2a})\Big] \,,\\
  \mathcal{T}^{(2a)}_2(t(\tau_{2a})) &\,=  \frac{(\tau_{2a}-3)}{\pi }
   \Big[-5 I(\eisb_{3,6,0,1}; \tau_{2a})+I(\eisb_{3,6,3,1}; \tau_{2a})+5
   I(\eisb_{3,6,3,2}; \tau_{2a})\Big] \\
   &+2 i (5
   I(1,\eisb_{3,6,0,1}; \tau_{2a})-I(1,\eisb_{3,6,3,1}; \tau_{2a})-5
   I(1,\eisb_{3,6,3,2}; \tau_{2a})) -\frac{i \pi }{8}\,,
  \esp\eeq
  
  \beq\bsp
\label{eq:finalresult-sunriseAB_2b}
 \mathcal{T}^{(2b)}_1(t(\tau_{2b})) &\,=  
 10 I(\eisa_{3,6,1,0},1; \tau_{2b})-2 I(\eisa_{3,6,1,3},1; \tau_{2b})-10
   I(\eisa_{3,6,2,3},1; \tau_{2b}) \\&-\frac{5 {\rm Cl}_2( \pi/3 )}{2 \pi } \,,  \\
  \mathcal{T}^{(2b)}_2(t(\tau_{2b})) &\,=  
  \frac{i}{\pi }
   \Big[5 I(\eisa_{3,6,1,0}; \tau_{2b})-I(\eisa_{3,6,1,3}; \tau_{2b})-5
   I(\eisa_{3,6,2,3}; \tau_{2b})\Big] \\
   &-30 I(\eisa_{3,6,1,0},1; \tau_{2b})+6
   I(\eisa_{3,6,1,3},1; \tau_{2b})+30 I(\eisa_{3,6,2,3},1; \tau_{2b}) \\
   &+\frac{15 {\rm Cl}_2( \pi/3 )-i \pi ^2}{2\pi }\,,
  \esp\eeq
  
  \beq\bsp
\label{eq:finalresult-sunriseAB_3a}
 \mathcal{T}^{(3a)}_1(t(\tau_{3a})) &\,=   -\frac{i (\tau_{3a}-1)}{\pi }
  \Big[5 I(\eisa_{3,6,1,0}; \tau_{3a})+5
   I(\eisa_{3,6,1,3}; \tau_{3a})+I(\eisa_{3,6,2,3}; \tau_{3a})\Big] \\
   &-10
   \Big[ I(1,\eisa_{3,6,1,0}; \tau_{3a})+I(1,\eisa_{3,6,1,3}; \tau_{3a})\Big]
   -2
   I(1,\eisa_{3,6,2,3}; \tau_{3a}) \\ &-\frac{5 {\rm Cl}_2( \pi/3 )}{2 \pi }\,, \\
  \mathcal{T}^{(3a)}_2(t(\tau_{3a})) &\,= \frac{i (3 \tau_{3a}-2)}{\pi } \Big[5 I(\eisa_{3,6,1,0}; \tau_{3a})+5
   I(\eisa_{3,6,1,3}; \tau_{3a})+I(\eisa_{3,6,2,3}; \tau_{3a})\Big] \\
   &+30
   \Big[ I(1,\eisa_{3,6,1,0}; \tau_{3a})+I(1,\eisa_{3,6,1,3}; \tau_{3a}) \Big]+6
   I(1,\eisa_{3,6,2,3}; \tau_{3a}) \\&+\frac{15 {\rm Cl}_2( \pi/3 )-i \pi ^2}{2 \pi }\,,
  \esp\eeq
  
  \beq\bsp
\label{eq:finalresult-sunriseAB_3b}
 \mathcal{T}^{(3b)}_1(t(\tau_{3b})) &\,=  \frac{(\tau_{3b}+1)}{\pi } \Big[-5 I(\eisb_{3,6,0,1}; \tau_{3b})-5
   I(\eisb_{3,6,3,1}; \tau_{3b})-I(\eisb_{3,6,3,2}; \tau_{3b}) \Big]\\
   &+2 i 
   \Big[5 I(1,\eisb_{3,6,0,1}; \tau_{3b})+5
   I(1,\eisb_{3,6,3,1}; \tau_{3b})+I(1,\eisb_{3,6,3,2}; \tau_{3b}) \Big] +\frac{i \pi }{8} \,,\\
  \mathcal{T}^{(3b)}_2(t(\tau_{3b})) &\,= \frac{(2 \tau_{3b}+3)}{\pi } \Big[5 I(\eisb_{3,6,0,1}; \tau_{3b})+5
   I(\eisb_{3,6,3,1}; \tau_{3b})+I(\eisb_{3,6,3,2}; \tau_{3b})\Big] \\
   &-4 i \Big[ 5
   I(1,\eisb_{3,6,0,1}; \tau_{3b})+5
   I(1,\eisb_{3,6,3,1}; \tau_{3b})+I(1,\eisb_{3,6,3,2}; \tau_{3b})\Big] -\frac{3 i \pi }{4}\,.
  \esp\eeq